\documentclass{aa}  
\usepackage{graphicx}
\usepackage{txfonts}
\usepackage{slashbox}
\usepackage{xcolor}

\usepackage{natbib}

\begin{document} 

   \title{Reconstructing solar magnetic fields from historical observations}

   \subtitle{VIII. AIA 1600~\AA{} contrast as a proxy of solar magnetic fields}

   \author{Ismo T\"ahtinen\inst{1}
          \and
                  I.I. Virtanen\inst{1,2}
          \and
          Alexei A. Pevtsov\inst{3}
          \and
          Kalevi Mursula\inst{1}
          }

   \institute{ReSoLVE Centre of Excellence, Space Physics and Astronomy research unit, University of Oulu, POB 8000, FI-90014, Oulu, Finland\\
              \email{ismo.tahtinen@oulu.fi};  \email{kalevi.mursula@oulu.fi}
         \and
              Oulu University of Applied Sciences, POB 222, FI-90101, Oulu, Finland\\
              \email{ilpo.virtanen@oamk.fi}
         \and
             National Solar Observatory, Boulder, CO 80303, USA\\
             \email{apevtsov@nso.edu}
         }

   \date{Received ; accepted}

  \abstract
   {The bright regions in the solar chromosphere and temperature minimum have a good spatial correspondence with regions of intense photospheric magnetic field.
Bright regions are visible in different emission lines and parts of the continuum.
Their observation started more than a hundred years ago with the invention of the spectroheliograph.
While the historical spectroheliograms are essential for studying the long-term variability of the Sun, the modern satellite-borne observations can help us reveal the nature of chromospheric brightenings in previously unattainable detail.  
   }
   {
Our aim is to improve the understanding of the relation between magnetic fields and radiative structures by studying modern seeing-free observations of far-ultraviolet (FUV) radiation around 1600~\AA{} and photospheric magnetic fields.
   }
   {
 We used Helioseismic and Magnetic Imager (HMI) observations of photospheric magnetic fields and Atmospheric Imaging Assembly (AIA) observations of FUV contrast around 1600~\AA{}.
We developed a robust method to find contrast thresholds defining bright and dark AIA 1600~\AA{} pixels, and we combine them to bright and dark clusters.
We investigate the relation of magnetic fields and AIA 1600~\AA{} radiation in bright and dark clusters.
   }
   {We find that the percentage of bright pixels (ranging from 2\% to 10\%) almost entirely explains the observed variability of 1600~\AA\ emission.
    We developed a multilinear regression model based on the percentages of bright and dark pixels, which can reliably predict the magnitude of the disk-averaged unsigned magnetic field.
    We find that bright and dark clusters closely correspond respectively to the populations of moderate (B~>~55 G) and strong (B~>~1365~G) magnetic field HMI clusters.
    The largest bright clusters have a constant mean unsigned magnetic field,  as  found previously for  Ca II K plages.
    However, the magnetic field strength of bright clusters is 254.7$\pm0.1$ G, which is roughly 100~G larger than found earlier for Ca II K plages. 
    
   }
   {}


   \maketitle

\section{Introduction}
\label{sec:intro}
The earliest observations of magnetic fields in astrophysics were sunspot field strength measurements at the Mount Wilson Observatory  (MWO) in California, USA, in 1908 \citep{Hale1908}.
Observations employed the Zeeman effect and were based on measuring the separation (splitting) between the two components of a spectral line, first \ion{Fe}{i}~6173 \AA\ and later \ion{Fe}{ii}~5250 \AA. 
Daily observations of sunspot magnetic fields have been conducted since 1917 \citep{Hale.etal1919}. 
In the early 1950s, the invention of an electronic magnetograph \citep{Babcock1953} allowed the measurement of regions with weaker magnetic fields than in sunspots, such as plages.
Regular full-disk magnetograms have been observed since early 1960, first at MWO \citep{Howard1974} and then at the National Solar Observatory (NSO) at Kitt Peak starting in 1973 \citep{Livingston.etal1976}.
In our long-term project aimed at reconstructing the past magnetic activity on the Sun \citep{Pevtsov.etal2016,Virtanen2017,Virtanen2018,Virtanen2019a,Pevtsov2019,Virtanen2019b}, we also use indirect measurements of magnetic fields obtained, in particular, from chromospheric spectroheliograms.
Recently,  \citet{Chatzistergos2021} and \citet{Shin2020} have used chromospheric observations to reconstruct the solar magnetic activity in the past.
Reconstructing past magnetic fields is important for understanding the long-term behavior of the Sun since it is the main factor affecting space climate.
Ca II K spectroheliograms are essential for this task, since the first ones were taken already in the 1890s in Europe \citep[Paris and Meudon Observatories, France;][]{Deslandres1909,Malherbe2019} and in the United States \citep[Kenwood Observatory;][]{Hale1893}.
Continuous observation campaigns in the Ca II K line started in the early 20th century in Kodaikanal (India), Mount Wilson Observatory (USA), the National Observatory of Japan (Japan), Paris-Meudon Observatory (France), the Arcetri Astrophysical Observatory (Italy), and the Astronomical Observatory of   Coimbra University (Portugal)
\citep[for a historical overview see, e.g.,][]{Bertello2016,Chatzistergos2020a}.
 Chromospheric spectroheliograms, together with flux transport simulations, have also been successfully used to study the evolution of large-scale solar magnetic fields, particularly the polar fields, which are only partially observable \citep{Virtanen2019a}.

A good spatial correspondence between bright chromospheric plages and areas of strong magnetic field called active regions (or magnetic plages) was already found in the early magnetographic observations
\citep{Babcock.Babcock1955,Leighton1959,Stepanov.Petrova1959}. 
The correlation between the intensity or the contrast (brightness after center-to-limb correction) of chromospheric brightenings and the unsigned magnetic flux density was confirmed in several studies \citep{Skumanich1975,Schrijver_1989,HarveyWhite1999,Ortiz_2005,Rezaei_2007,Loukitcheva2009,Kahil_2017,Kahil2019,Chatzistergos2019b}.
\citet{Loukitcheva2009} and \citet{Barczynski2018} confirmed this correlation also for 1600~\AA{} emission and magnetic flux density.
This relation is usually modeled as a power law, following \citet{Schrijver_1989} who found it to be a good description between the Ca K II line core excess brightness and the absolute value of the magnetic flux density.
\citet{Schrijver_1989} found the power-law exponent to be 0.6, but in later studies it  varied between 0.2 and 0.7, depending on the activity of the regions included.
Studies that  include active regions usually find a higher exponent than studies of a more quiet Sun.
\citet{Rezaei_2007} found that the exponent depends strongly on a lower cutoff of magnetic flux density.

Quite remarkably, \citet{Schrijver_1989} found that the  power-law relation with exponent 0.6 also exists between the Ca K emission and magnetic flux density of Sun-like stars.
A recent review of pixel-by-pixel studies can be found in \citet{Chatzistergos2019b}.
In addition to Ca II H \&~K, many other spectral lines and some parts of the continuum spectrum have also been related to the photospheric magnetic fields \citep{Fludra2002,Loukitcheva2009,Barczynski2018,Yeo2019}.
The correlation between the magnetic field and emission brightness typically decreases with emission height \citep{Barczynski2018}.
The far-ultraviolet (FUV) continuum from 1220 \AA{} to 2000 \AA{} is an interesting region since it primarily forms around the temperature minimum and the low chromosphere.
Due to its low emission height, FUV radiation is closely related to photospheric magnetic fields.
The Transition Region and Coronal Explorer (TRACE) was the first instrument to observe FUV radiation in detail with its three channels of 1550 \AA{}, 1600~\AA{}, and 1700 \AA{} \citep{Handy1999}.
 \citet{Rutten1999} demonstrated that bright FUV features in the quiet Sun closely correspond to features observed in Ca II K.
\citet{Loukitcheva2009} showed that the intensity of the TRACE 1600~\AA{} radiation is linearly proportional to Ca II K emission.
The Atmospheric Imaging Assembly \citep[AIA,][]{Lemen.etal2012} on board the Solar Dynamics Observatory \citep[SDO,][]{Pesnell.etal2012} continues to observe the FUV continuum on two channels, 1600~\AA{} and 1700 \AA{}.

Despite numerous pixel-by-pixel studies, surprisingly few studies between ensemble or disk-averaged quantities of radiative structures and magnetic fields exist.
One such study is by \citet{Schrijver1987}, who constructed several relations between parameters that characterize large-scale properties of active regions.
A most remarkable result of this study was that the magnetic flux contained within a plage linearly scales with the area of plage (i.e., the magnetic flux density is roughly constant over a plage).
He found an average unsigned flux density of B~$=~100~\pm~20$~G for a plage.
This scaling between the magnetic flux and the area of a plage has been employed by \citet{Pevtsov.etal2016} and \citet{Virtanen2019a} to reconstruct past magnetic fields from synoptic Ca II K maps.
The advantage of studying ensembles like plages instead of individual pixels is that ensembles are more likely to provide more robust results, although naturally pixel-by-pixel  reconstructions can in principle give more detailed information on the structure of plages.
The challenge that ensemble-based methods face is that there is no consensus on how to classify different structures.
In addition to ensemble averages, disk-averaged quantities are   of great interest since they provide proxies of the total flux of the solar magnetic field, which is one of the most important quantities quantifying the solar magnetic field.
Moreover, disk-averaged quantities of suitable spectral lines and continuum bands can be used as a proxy of the magnetic field of cool stars \citep{Reiners2012}.

Here we explore the relationship between radiative structures and magnetic fields using modern atmospheric seeing-free observations obtained by the Atmospheric Imaging Assembly \citep[AIA,][]{Lemen.etal2012} and Helioseismic Magnetic Imager \citep[HMI,][]{Scherrer.etal2012} on board the Solar Dynamics Observatory 
\citep[SDO,][]{Pesnell.etal2012}.
Instead of focusing on individual pixels, we mostly concentrate here on disk averages and average properties over specific ensembles to be called clusters.
Clusters may include, but do not distinguish between structures like plages, network, internetwork grains.
Two different types of clusters are defined based on two thresholds, one for dark and another for bright clusters.
We define these thresholds in an objective way using only statistics and with no visual feedback on how the clusters look.

This paper is organized as follows. 
In Sect. \ref{sec:data} we present the data, and in Sect. \ref{sec:calibration} the calibration of observations.
In Sect. \ref{sec:thresholds} we define the bright and dark pixels and respective clusters.
In Sect. \ref{sec:disc_int} we present our results for disk-integrated quantities,  and in Sect. \ref{sec:clusters} for cluster-integrated quantities.
In Sect. \ref{sec:cak} we discuss the relation between  AIA 1600~\AA{} and Ca II K emission.
We discuss our results in Sect. \ref{sec:disc} and give our conclusions in Sect. \ref{sec:conc}.

\section{Data}
\label{sec:data}

We use observations from the Atmospheric Imaging Assembly \citep[AIA,][]{Lemen.etal2012} and the Helioseismic and Magnetic Imager \citep[HMI,][]{Scherrer.etal2012} on board the Solar Dynamics Observatory 
\citep[SDO,][]{Pesnell.etal2012}. 
We use HMI 720-second data, which are full-disk line-of-sight (LOS) magnetograms computed every 12 minutes from filtergrams of the vector-field camera.
We prefer HMI 720-second data over HMI 45-second data since it has a lower noise level and since we use only one magnetogram per day.
The spatial resolution of HMI data is 1 arcsecond (0.5 arcsecond per pixel).

We employ AIA 1600~\AA{} emission to identify chromospheric brightenings. The
AIA data are full-disk images taken at 12-second cadence and with a spatial resolution of 1.2 arcsecond (0.6 arcsecond per pixel).
This wavelength band represents the FUV  continuum with some contribution from C \rm{IV} emission from the transition region.

Level 1 HMI and AIA data are available at the Joint Science Operations Center (JSOC\footnote{\url{http://jsoc.stanford.edu/}}).
These data differ in spatial resolution, in the location of the disk center, and in the angle between the vertical axis and solar rotation axis (p-angle).
However, these differences can be unified with the  JSOC data export tool before downloading.
We used this data export tool to co-align the HMI and AIA images and to unify them to the same 0.6 arcsec/pixel scale.
This also corresponds to the standard AIA level 1.5 processing.
The size of both HMI and AIA images is 4096 x 4096 pixels.
During our analysis we noticed that AIA and HMI images display small (1--2 pixel) co-alignment offsets.
We corrected for this by computing optimal shifts based on the Jaccard distance \citep{Bandyopadhyay2013} between HMI and AIA masks defined in Sect. \ref{sec:thresholds}.
 The computed offsets remained within $\pm$2 pxl in the x- and/or y-direction.
The offsets we found are consistent with earlier studies \citep{Orange2014,Yeo2019}.

Given the synoptic nature of our project, for each day of observations only one pair of HMI-AIA images was selected and analyzed.
Using one single HMI magnetogram per day taken at about the same Universal Time (5:48) also minimizes the 
variation in magnetic field strength due to SDO orbital motion, which can be up to 5\% with 12-hour periodicity \citep[][]{Pevtsov2016,Smirnova.etal2013}. 
Since a high radial velocity with respect to the Sun, together with strong magnetic fields, can lead to inaccuracies in magnetic field measurements, we selected a time that minimizes the average radial velocity of the instrument and maximizes the number of observations \citep{Liu_2012}.
On average, SDO has the smallest radial velocity of about 400 m/s at local midnight (7 UT) and noon (19 UT). 
However, many observations are missing during these times of the day.
In order to balance between these restrictions, we chose the time at 05:48:00 UTC, when the average radial velocity of the instrument is still relatively slow, about 800 m/s, compared to 1 UT or 13 UT when SDO radial speed is 2500 m/s.
Data were missing on 11 days, and we found problems with data quality on 21 days, which we removed from the final dataset.
Finally, we transformed the line-of-sight $B_{LOS}$ field to radial field $B_{r}~=~B_{LOS}/\mu$, where $\mu$~=~cos$\theta$ is the cosine of the heliocentric angle $\theta$.
The AIA images   selected were  those that are closest in time to the HMI observations.
The time difference between the AIA and HMI observations is within one minute.
Our final dataset consists of 2509 AIA and HMI images at 24-hour cadence measured at 05:48:00 UTC from 13 May 2010 to 9 June 2017.
A more thorough analysis was performed for 1165 images from 1 March 2014 to 9 June 2017.
In Sect. \ref{sec:disc_int} these 1165 images were divided into two sets of 765 and 400 images, due to a change in the HMI measurement scheme on 13 April 2016 \citep{Liu.etal2016}.

\section{Calibration of AIA and HMI observations}
\label{sec:calibration}
Our aim for this paper was to compare HMI and AIA observations using the selected dataset, which covers several years, and therefore requires that each dataset has a stable calibration.
While the HMI data has no major issues or anomalies related to the instrument, there is a question regarding the  AIA absolute photometry. 
The degradation of AIA detectors affects the results of the long-term variation of AIA 1600~\AA{} brightness. 
The most significant instrumental defects are the nonuniformities  in camera detector cells and the gradual overall degradation of detectors.
A nonuniformity is corrected by flat-fielding, which means that the observed value in each cell is divided by a corresponding correction factor, which normalizes the observed brightness in that pixel.
Flat-field matrices are derived from observations and, in the case of AIA standard data processing, are updated approximately every three months.
This flat-fielding processing has been done at JSOC on the level 1 data that we use.

\begin{figure*}
\centering
\includegraphics[width=\textwidth,height=\textheight,keepaspectratio]{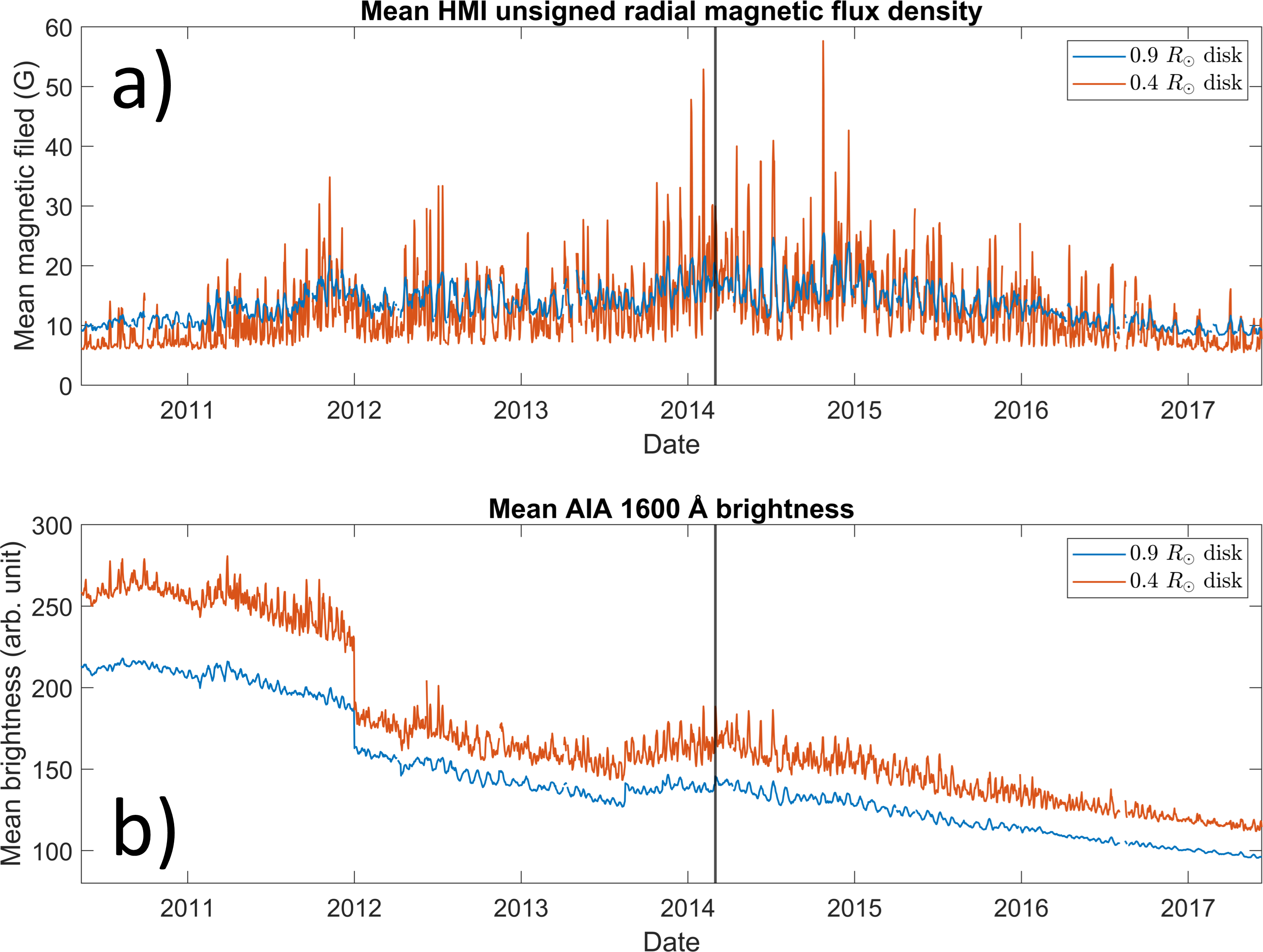}
   \caption{Timeseries of HMI and AIA 1600~\AA{} observations. a) Mean unsigned magnetic field and b) mean AIA 1600~\AA{} brightness averaged over a $0.9R_{\sun}$-radius disk (blue) and a $0.4R_{\sun}$-radius disk (orange) from 13 May 2010 to 9 June 2017. The black vertical line at March 2014 shows the starting point of the final dataset used. \label{HMI_AIA_timeseries}}
\end{figure*}

Figure \ref{HMI_AIA_timeseries}a shows the mean HMI unsigned magnetic field averaged over a $0.9R_{\sun}$-radius disk and over a $0.4R_{\sun}$-radius disk from 13 May 2010 to 9 June 2017.
The mean unsigned magnetic field is seen to vary, roughly in phase with solar activity over solar cycle 24. 
The short-term fluctuations relate to solar rotation and the uneven longitudinal distribution of solar activity. 
Variability is greater for the more limited $0.4R_{\sun}$-radius disk mainly because the relative contribution of active regions increases when the polar regions are excluded.
The rotation minima of the mean unsigned magnetic field are slightly higher for the larger disk since we  transformed LOS values to radial and since the noise level increases towards the limb. 

Figure \ref{HMI_AIA_timeseries}b shows the mean AIA 1600~\AA{} brightness averaged over the $0.9R_{\sun}$-radius disk and over the $0.4R_{\sun}$-radius disk from 13 May 2010 to 9 June 2017.
AIA brightness does not show a consistent solar cycle variation, but rather a long-term declining trend due to the degradation of detectors.  
In addition, two significant discontinuities are seen in the  AIA data, one (decreasing step) at the start of 2012 and the other (increasing step) in August 2013.
There are also smaller discontinuities, for example  in November 2013 and March 2014.
According to the AIA team, all these changes are due to changes in the flat-fields. 
At the beginning of the mission a dummy flat-field was used, but since 2012 the flat-fields have been based on observed values.
The AIA team  has informed us that there are altogether 13 discontinuities in the mean brightness data due to changes in flat-fields (private communication with Wei Liu, 2017). 
Small steps in mean brightness occur during every flat-field change, but they are too small to be noticed from the daily data in Fig. \ref{HMI_AIA_timeseries}b.

\begin{figure*}
\centering
\includegraphics[width=\textwidth,height=\textheight,keepaspectratio]{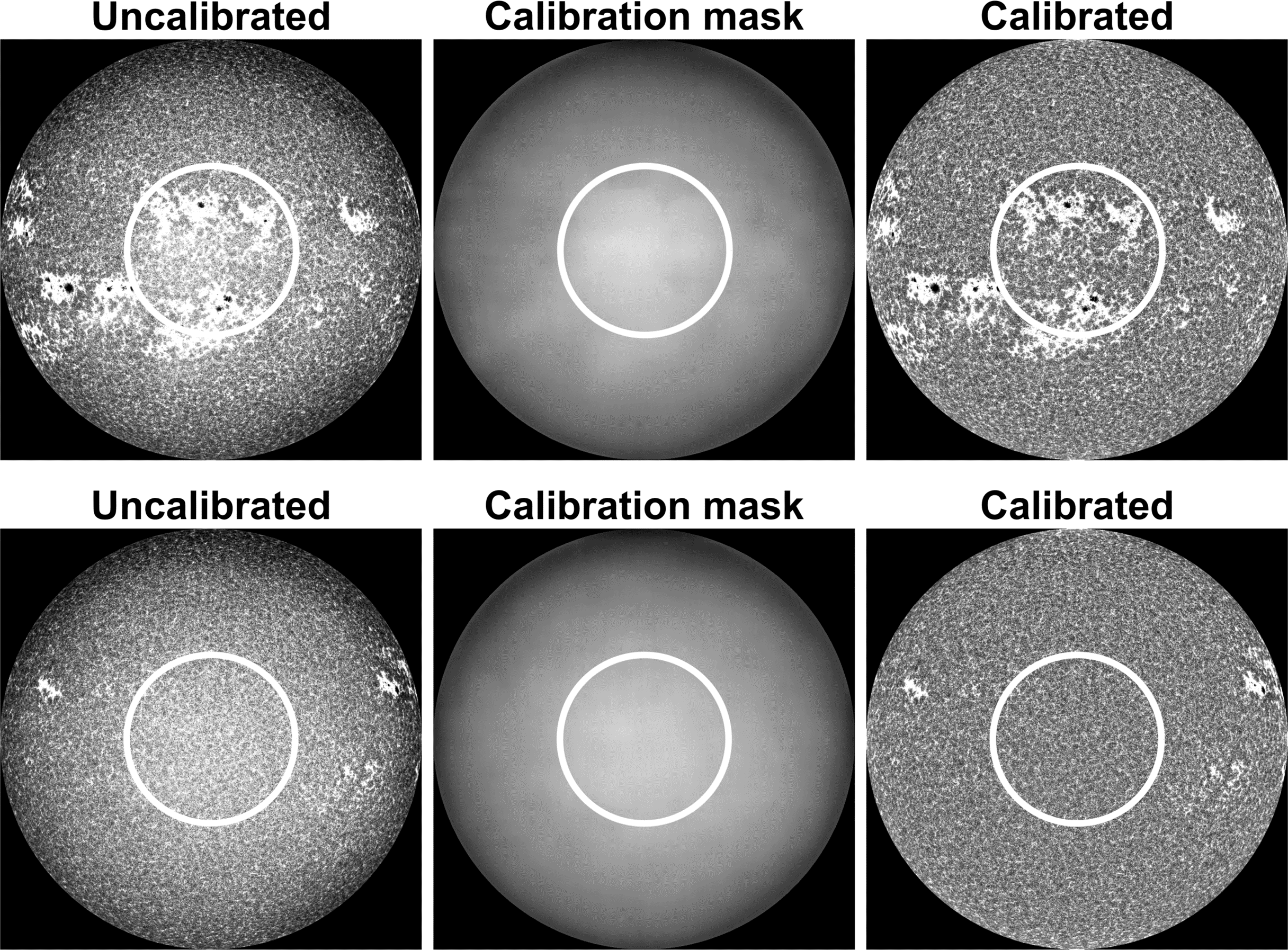}
\caption{Examples of image calibration for an active day (5 July 2014, upper row) and a quiet day (28 May 2017, lower row) images. First column: Uncalibrated images. Second column: Calibration masks representing the solar background. Third column: Calibrated images. White circles show the region within $0.4R_{\sun}$ from the disk center. Color range of pixels is 0.5 -- 1.5. Uncalibrated images and calibration masks were normalized by their median brightness.}
\label{Calibration}
\end{figure*}

In order to have a reliable long-term evolution of AIA 1600~\AA{} brightness, we needed to correct both for the instrument degradation and the flat-field discontinuities.
We first produced a quiet-Sun image by removing all AIA pixels whose LOS magnetic flux density is above 10~G in corresponding HMI image.
We then calculated the center-to-limb (CTL) profile by fitting a fifth-order polynomial in $\mu$ \citep{Neckel1994} to the remaining pixels.
We then removed the CTL variation from the quiet-Sun image by dividing it with the CTL profile image.
Finally, we applied a R/8 median filter to this normalized quiet-Sun image.
Before median filtering we set all the missing pixel values to I~=~1.
We also downscaled the normalized quiet-Sun image with a factor of 1/8 before applying median filter and then resampled the filtered image back to the original resolution to reduce computation time \citep{Bose2018}.
The final calibration mask was produced by multiplying median filtered quiet-Sun images with the CTL profile image.
The original AIA image was calibrated by dividing it with this calibration mask.
This calibration process was applied to all images, and it normalizes the pixel values of images to the momentary background level.
We selected the size of median filter by studying the distribution of pixel values of the average quiet Sun (averaged over all calibrated quiet-Sun images) with different filter sizes.
We selected R/8 since it minimizes the kurtosis (K~=~3.179) of the average quiet-Sun distribution.
This value is close to the kurtosis (K~=~3) of the normal distribution, which the contrast distribution of the average quiet Sun should resemble.
We note that R/6 produces an almost  equal result (K~=~3.180).
These results are in agreement with \citet{Chatzistergos2018} who found R/6 and R/8 to be the best values for median filtering Rome/PSPT Ca II~K data.
This calibration method is partly based on the one used by \citet{Bose2018} to remove the limb darkening from AIA 1600~\AA{} data.
However, they did not take into account the bright regions, and used a larger (approximately R/2) median filter.
The average contrast of calibrated AIA images over the $0.9R_{\sun}$ ($0.4R_{\sun}$) disk is I~=~1.07 (1.10 respectively). 
We note that these means are somewhat larger than 1 because of the asymmetric, non-Gaussian, distribution of contrasts (see later in Sect. \ref{sec:disc_int}).
The ratio of brightness to background/CTL variation is generally called contrast in the literature, and we use it from now on when referring to the brightness of calibrated AIA images.
Figure \ref{Calibration} shows the effect of calibration for an active day (upper row) and a quiet day (lower row) image.
The white circles indicate the region of $0.4R_{\sun}$ from the disk center, which we  mostly use in the  analysis later on.

We tested the validity of this calibration by producing the photometric sum series \citep{Preminger2002,Chatzistergos2020b} from the calibrated full-disk images.
The photometric sum is a sum over a contrast image whose CTL variation has been restored

\begin{equation}
    \Sigma~=~\sum_{i} \mathrm{(I_i-1)}*\mathrm{CTL_i},  
\end{equation}
where $\mathrm{I_i}$ are the pixel values in calibrated AIA contrast image, $\mathrm{CTL_i}$ are the pixel values in normalized (I~=~1, at the disk center) CTL profile image, and the sum runs through all the pixels.
The photometric sum for the average quiet Sun equals zero.
We compared photometric sums series against the expected radiation that AIA 1600~\AA{} would measure without degradation.
The AIA 1600~\AA{} brightness consists of radiation of wavelengths with 97 \% of contribution coming from the range  1400--1700 \AA{}.
The wavelength dependence is captured by the response function presented in Fig. 9 of \citet{Boerner2012}.
We derived the expected radiation from the AIA 1600~\AA{} response function using the measurements of the Solar Stellar Irradiance Comparison Experiment (SOLSTICE) on board the Solar Radiation and Climate Experiment (SORCE).
SOLSTICE measures the spectral irradiance at wavelengths from 1150 \AA{} to 3100 \AA{} with a spectral resolution of 1 \AA{} \citep{McClintock2005}, thus also covering the whole range of wavelengths contributing to the AIA 1600~\AA{} measurements.
SOLSTICE has operated since 2003, but, due to a battery anomaly, there is a gap in observations from 13 July 2013 to 25 February 2014.

\begin{figure*}
   \centering
  \includegraphics[width=\textwidth,height=\textheight,keepaspectratio]{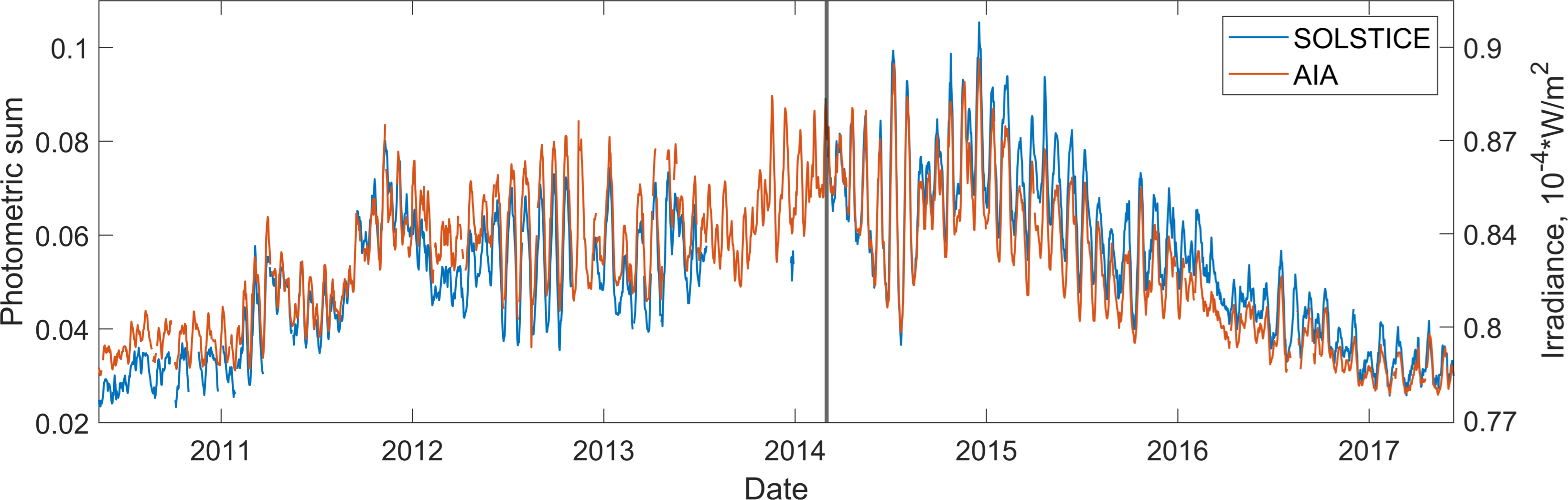}
   \caption{Daily photometric sums of calibrated AIA 1600~\AA{} images (orange) and the expected irradiance from correlated SOLSTICE observations (blue) from 1 March 2014 to 9 June 2017. Black vertical line on March, 2014 shows the starting point of our final dataset used. \label{CalibratedAll}}%
    \end{figure*}

\begin{figure*}
   \centering
  \includegraphics[width=\textwidth,height=\textheight,keepaspectratio]{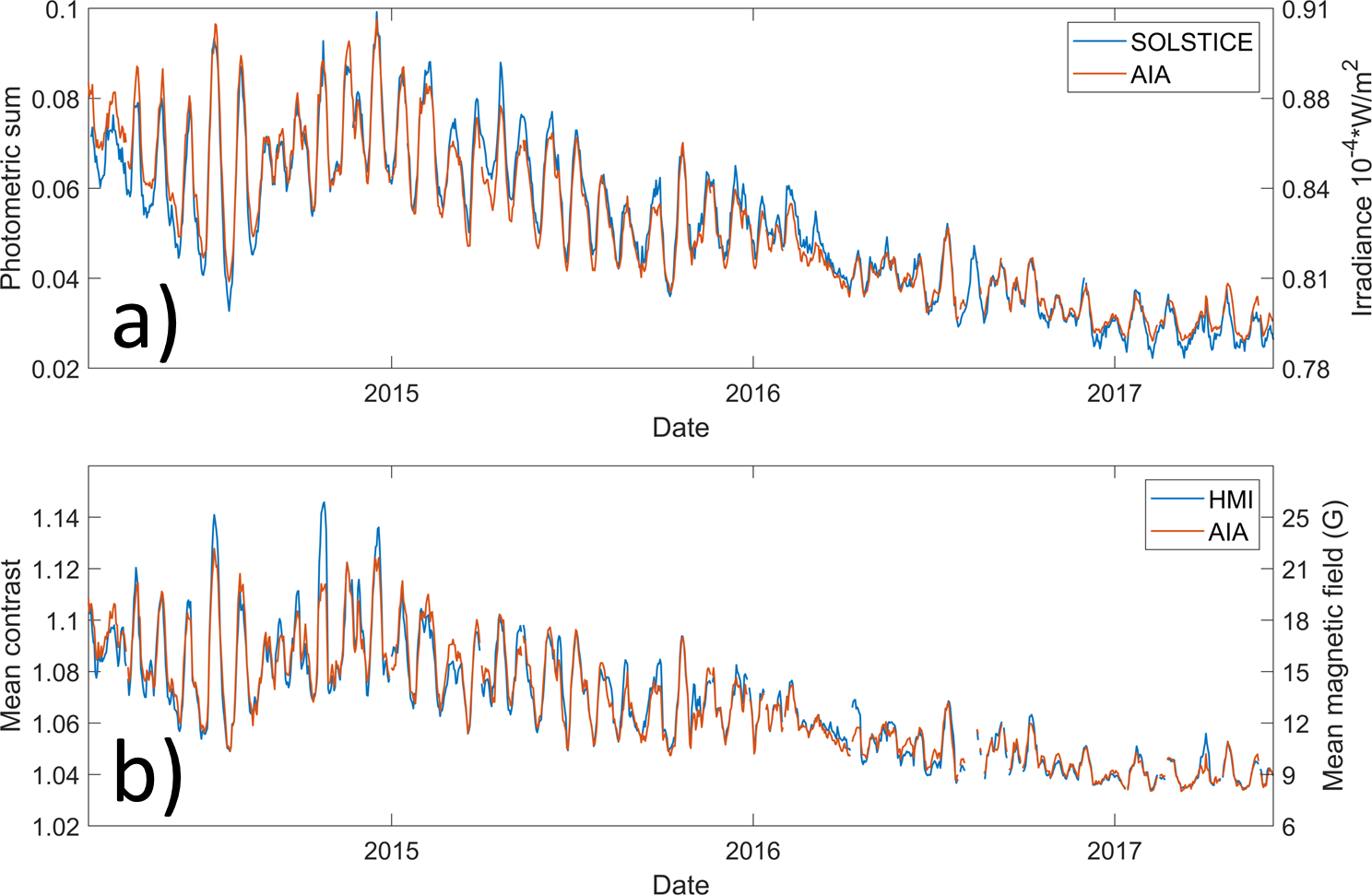}
   \caption{Timeseries from 1 March 2014 to 9 June 2017. a) AIA 1600~\AA{} photometric sum (orange, left axis, arb. units) and expected irradiance from correlated SOLSTICE observations (blue, right axis, arb. units). b) AIA 1600~\AA{} mean contrast (orange, left axis) and HMI mean unsigned magnetic field (blue, right axis, G) within $0.9R_{\sun}$. \label{CalibratedEnd}}%
    \end{figure*}

Figure \ref{CalibratedAll} shows the photometric sum series of calibrated AIA 1600~\AA{} full-disk images together with the SOLSTICE irradiance weighted by the AIA 1600~\AA{} response function.
The photometric sum series and SOLSTICE data are presented in AIA scale.
This correlated SOLSTICE irradiance represents the intensity of the emission that we would expect AIA 1600~\AA{} to measure without degradation.
Figure \ref{CalibratedAll} shows that the photometric sum series varies quite similarly to the SOLSTICE observations, demonstrating that the adopted calibration procedure corrects for the instrument degradation.
Calibration corrects the dramatic declining trend seen in Fig. \ref{HMI_AIA_timeseries}b and restores the solar cycle variation of the  AIA 1600~\AA{} contrast (also seen in SOLSTICE measurements).
However, Fig. \ref{CalibratedAll} shows that the difference between the  AIA photometric sum series and the SOLSTICE irradiance was somewhat larger in the beginning of the time interval before 2012, when the first proper flat-fieldings were introduced to the AIA data.
The higher SOLSTICE values in 2015--2016 are due to the correlation of data for the whole interval, and this difference vanishes in Fig. \ref{CalibratedEnd}.

As already noted above, the last significant jump in AIA 1600~\AA{} mean brightness happened at the start of March 2014, which is denoted by the vertical black line in Fig. \ref{HMI_AIA_timeseries} and Fig. \ref{CalibratedAll}.
SOLSTICE restarted observations after the battery anomaly just a few days earlier, on 25 February.
We limit the study period to March 2014 onward, during which time the AIA 1600~\AA{} photometric sum series variability is very close to the SOLSTICE variability, as depicted in Fig. \ref{CalibratedEnd}a (where we re-correlated the SOLSTICE data with AIA).
 The first proper flat-fielding started in 2012, and the quality of flat-fielding has improved since then.
Furthermore, it is likely that the need for these corrections has become less urgent with the decreasing irradiance in the declining phase of a solar cycle.

Accordingly, we limit our final dataset to this period from 1 March 2014 to 9 June 2017.
The daily mean values of both AIA and HMI for this period are shown in Fig. \ref{CalibratedEnd}.
Moreover, although the calibration mostly corrects for the limb darkening, we  decided to limit most of our analysis to within $0.4R_{\sun}$ from the disk center, which includes most active regions in this period including the maximum and the declining phase of cycle 24.
Using a smaller disk also has the benefit that it enhances the dynamic range of the observed activity by leaving out the less active polar regions.
Naturally, some nonuniformities still remain even within the smaller $0.4R_{\sun}$ disk, but their effect is smaller there than closer to the limb.

\section{Definition of bright and dark pixels and clusters and AIA-HMI comparison}
\label{sec:thresholds}

Increased magnetic activity can either increase or decrease the AIA 1600~\AA{} contrast.
Canonical examples are plages and sunspots,  both of which can be seen in the  upper row  of Fig. \ref{Calibration}.
Here we define two thresholds that we  use to separate these two distinct contrast classes from the more moderate radiation.
Pixels with contrast below the lower threshold are called dark pixels, and pixels exceeding the higher threshold are called bright pixels.
Furthermore, we define bright and dark clusters as 4-connected regions of either bright or dark pixels.
The term 4-connected indicates that every pixel in a cluster is connected to others by at least one of its four sides.
Bright and dark clusters are used here as generic names for bright structures, such as plages and network, and dark structures, such as sunspots and pores.

\begin{figure*}
\centering
\includegraphics[width=\textwidth,height=\textheight,keepaspectratio]{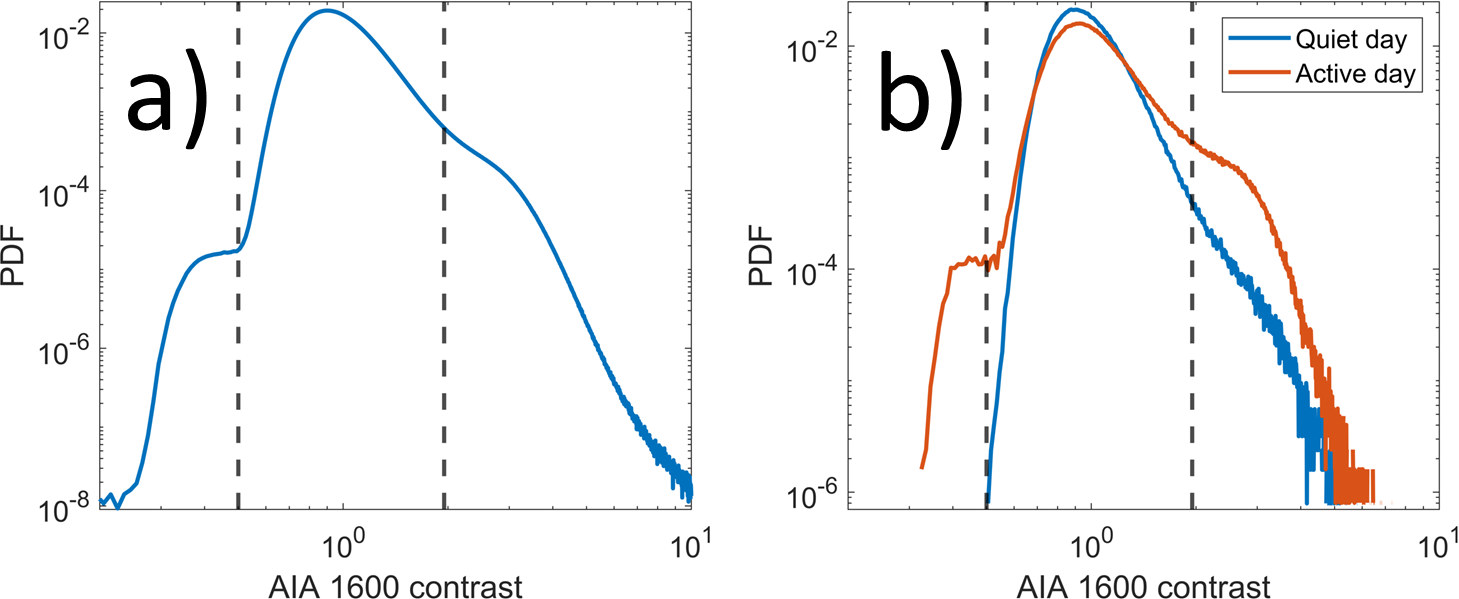}
\caption{Distribution of AIA 1600~\AA{} pixel contrasts within $0.4 R_{\sun}$ disk. a) All images from 1 March 2014 to 9 June 2017. b) A quiet day (28 May 2017, blue line) and an active day (5 July 2014, orange line). Left dashed line: Dark pixel threshold (I~=~0.5). Right dashed line: Bright pixel threshold (I~=~1.95).}
\label{fig:Histograms}%
\end{figure*}

Panel a of Fig. \ref{fig:Histograms} shows the probability density function (PDF) of AIA 1600~\AA{} pixel contrast for all the AIA 1600~\AA{} images, whereas panel b shows the contrast PDF distributions for the typical active (5 July 2014) and quiet (28 May 2017) day of Fig. \ref{Calibration}.
Panel b of Fig. \ref{fig:Histograms} shows a notable difference between an active-day and a quiet-day contrast distribution.
On a quiet day the distribution is narrower compared to an active day.
The most notable difference is active-day the low-contrast pixel population below I~<~0.5 (we use here I to denote AIA 1600~\AA{} contrast), which is nonexistent on a quiet day.
The second difference is that, although both the active-day and quiet-day distributions are right-skewed, there is a much more noticeable bump on the active-day distribution.
Based on the clear difference between the active-day and quiet-day PDF distributions below I~<~0.5, we chose $I_{\mathrm{DP}}~=~0.5$ as an upper threshold (left dashed line in Fig. \ref{fig:Histograms}) for the population of dark pixels.
However, it is unclear from Fig. \ref{fig:Histograms} what exactly would be a suitable threshold value to distinguish the bright pixels from the moderate background pixels.

\subsection{Bright pixels and clusters}
\begin{figure*}
 \centering
 \includegraphics[width=\textwidth,height=\textheight,keepaspectratio]{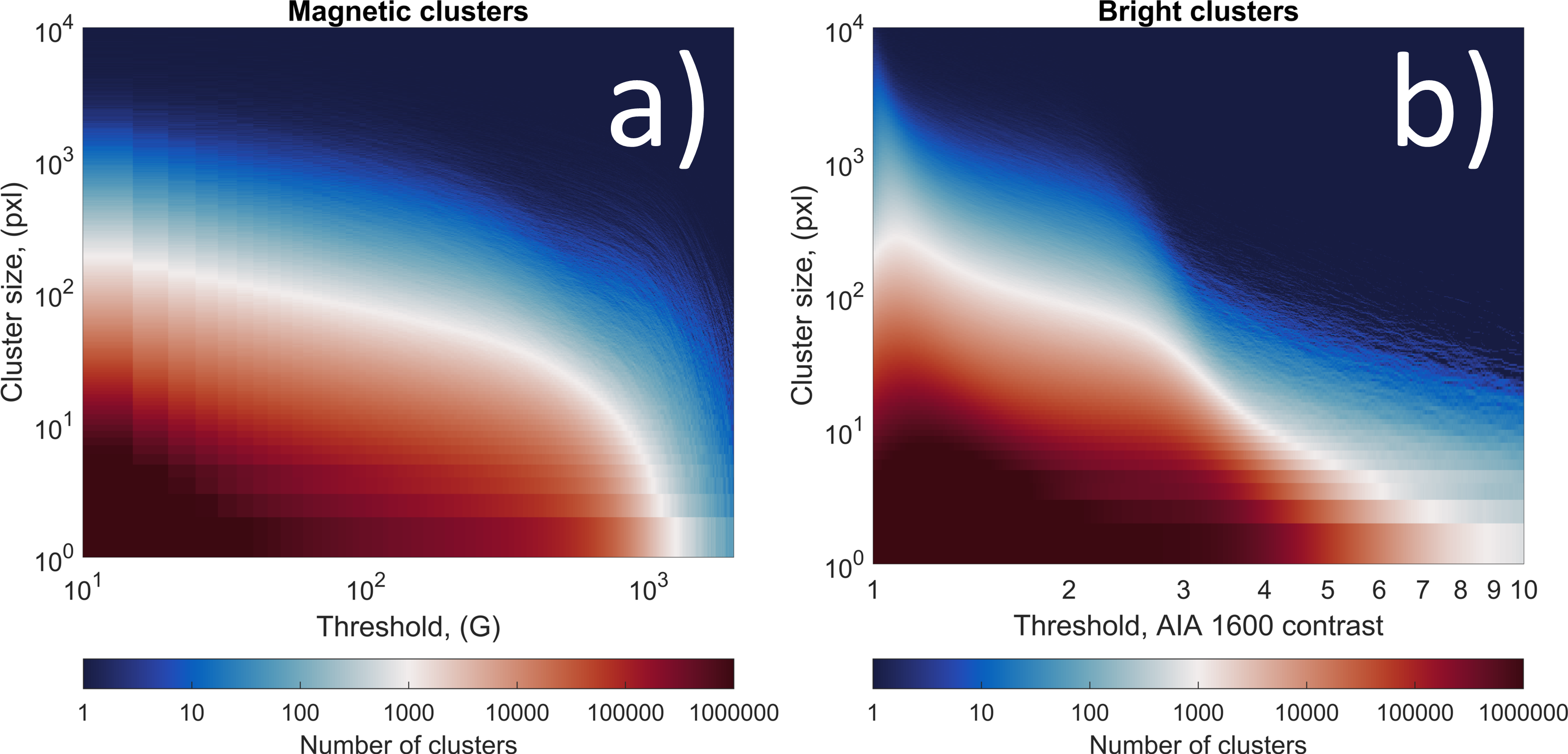}
 \caption{Size distributions of magnetic (panel a) and bright (panel b) clusters for different thresholds in all images from 1 March 2014 to 9 June 2017.
 \label{fig:NumberOfClusters}}
\end{figure*}

\begin{figure*}
 \centering
 \includegraphics[width=\textwidth,height=\textheight,keepaspectratio]{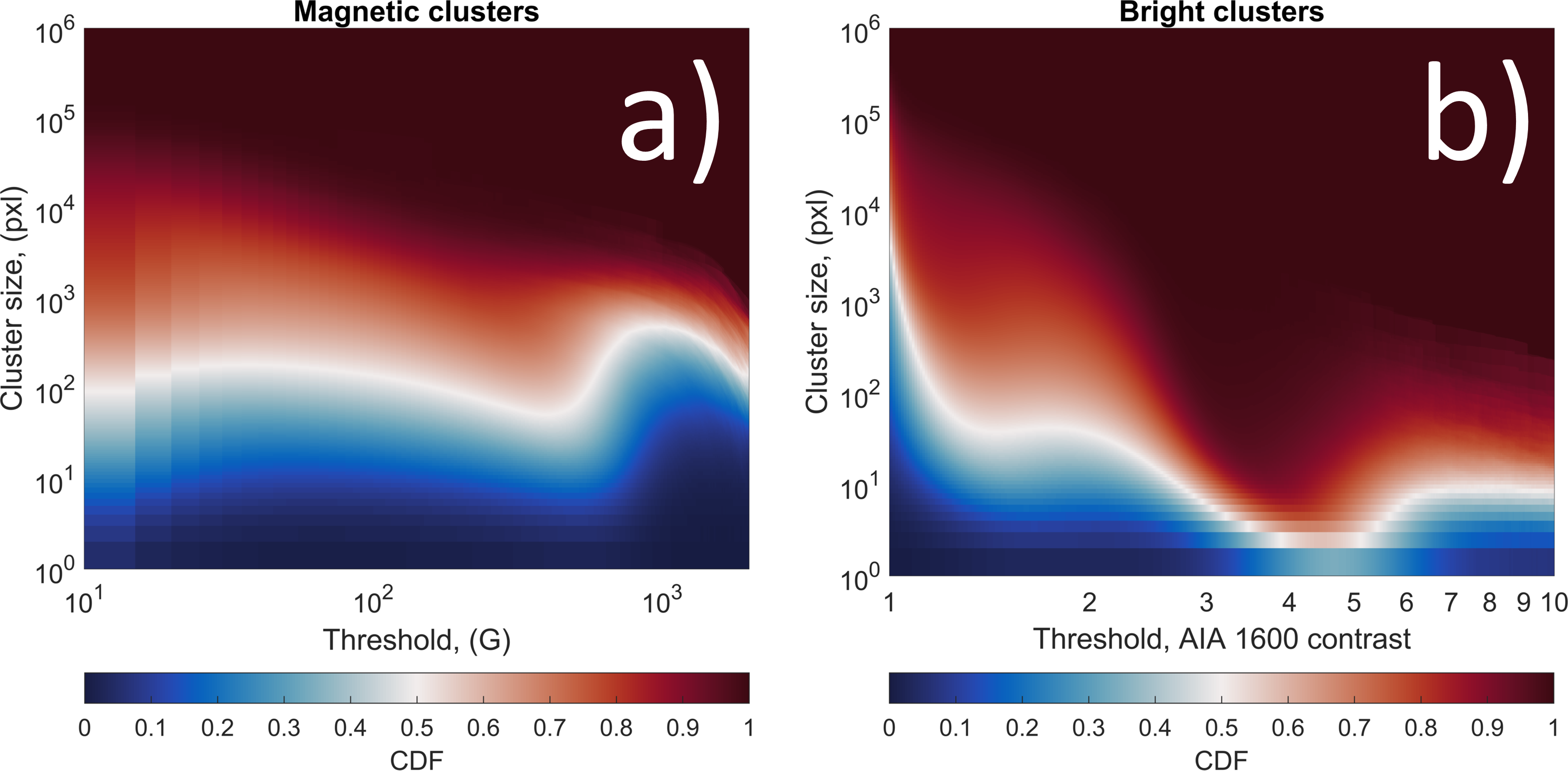}
 \caption{Size weighted cumulative distributions of magnetic (panel a) and bright (panel b)  cluster sizes for different thresholds in all images from 1 March 2014 to 9 June 2017.
 \label{fig:ClusterCumDist}}
\end{figure*}

To find a suitable threshold for the bright pixels, we studied how the number and size of bright AIA and magnetic HMI clusters behave as a function of threshold.
Clusters are defined as 4-connected regions of HMI/AIA pixels that exceed a given threshold (absolute value of HMI field/AIA 1600 \AA{} contrast).
Figure \ref{fig:NumberOfClusters} shows a two-dimensional histogram of magnetic (panel a) and bright (panel b) clusters.
The horizontal axis shows the applied threshold, and the vertical axis the cluster size in pixels.
The range of magnetic thresholds for HMI clusters is from 10~G to 2000~G with steps of 5~G, while the range of contrast thresholds for AIA clusters is from 1 to 10 with steps of 0.01.
The color shows the number of clusters for a given threshold and cluster size.
We note that there are larger clusters than A~=~$10^4$ pxl, but we do not show them in Fig. \ref{fig:NumberOfClusters}, since they can not be distinguished from the plot due to their small number.
The presence of larger clusters can be seen in Fig. \ref{fig:ClusterCumDist}, which shows weighted cumulative distribution functions (CDFs) for different threshold values for both HMI magnetic clusters (panel a) and AIA bright clusters (panel b).
We have weighted the distributions by the cluster size.
The color shows the fraction of pixels above the threshold that belong to smaller clusters than size A.
Figure \ref{fig:ClusterCumDist}a shows CDFs calculated for magnetic clusters in steps of 5~G from 10~G to 2000 G.
These CDFs display a bimodal structure, which can be seen from how the  color white, which  corresponds to the median, shifts at different threshold values.
The fraction of pixels in large clusters first grows with increasing threshold, but their number starts to decrease around 50~G.
This decrease continues until about 500~G when the number of pixels again starts to grow with the threshold.
The fraction of pixels that belong to large clusters is highest at around 1000 G.
However, as can be seen from Fig. \ref{fig:NumberOfClusters}a, the number of clusters above 1000~G is quite small in comparison to smaller thresholds.
Figure \ref{fig:ClusterCumDist}b shows CDFs calculated for bright clusters in steps of 0.01 from 1 to 10.
With a threshold of I~=~1, the fraction of bright pixels belonging to large clusters is highest.
However, the number of pixels belonging to large clusters rapidly decreases by over two orders of magnitude before flattening around I~=~1.4.
Above I~=~2, the number of pixels in large clusters starts to decrease again, and at around I~=~4 median has decreased to A~=~2 pxl meaning that 50\% of the pixels above this threshold belong to clusters of  1 or 2 pixels in size.
Above I~=~5, the number of pixels in larger clusters again increases and then flattens.
Again, the number of clusters above I~=~5 is quite small compared to smaller thresholds.
There are days without a single pixel above I~=~6.28, so these clusters represent sporadic events not always present within $0.4R_{\sun}$ solar disk.

\begin{figure*}
 \centering
 \includegraphics[width=\textwidth,height=\textheight,keepaspectratio]{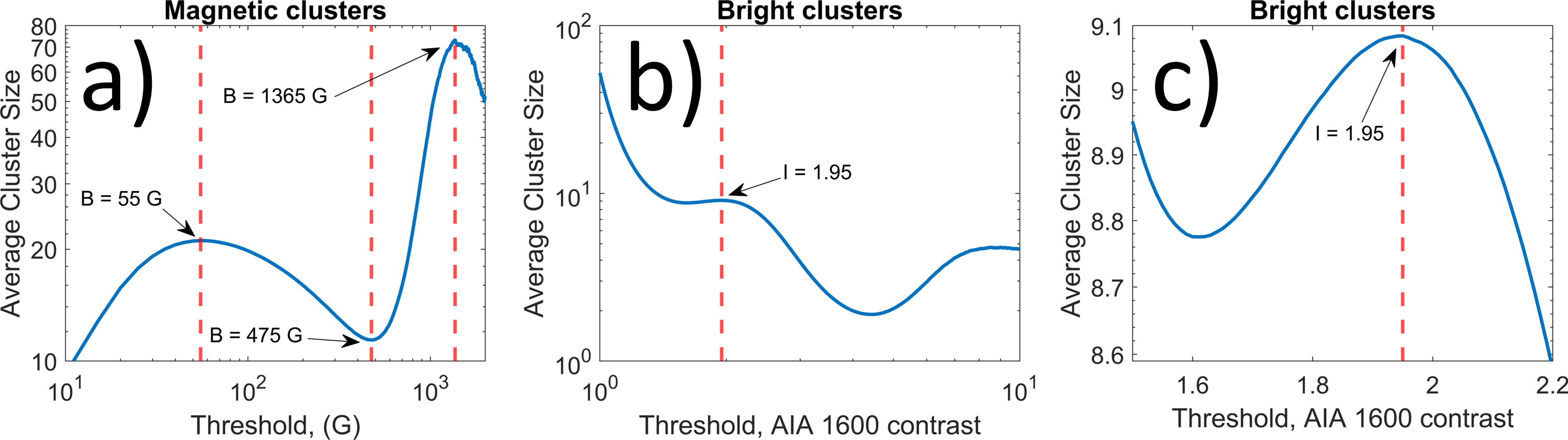}
 \caption{Average clusters sizes for different thresholds in all images from 1 March 2014 to 9 June 2017. a) Magnetic clusters. The red lines show two local maxima and one local minimum. b) Bright clusters. The red line shows a local maximum. c) Same as panel b, but with limited x-axis range.
 \label{fig:ClusterDistributions}}
\end{figure*}

\begin{figure*}
 \centering
 \includegraphics[width=\textwidth,height=\textheight,keepaspectratio]{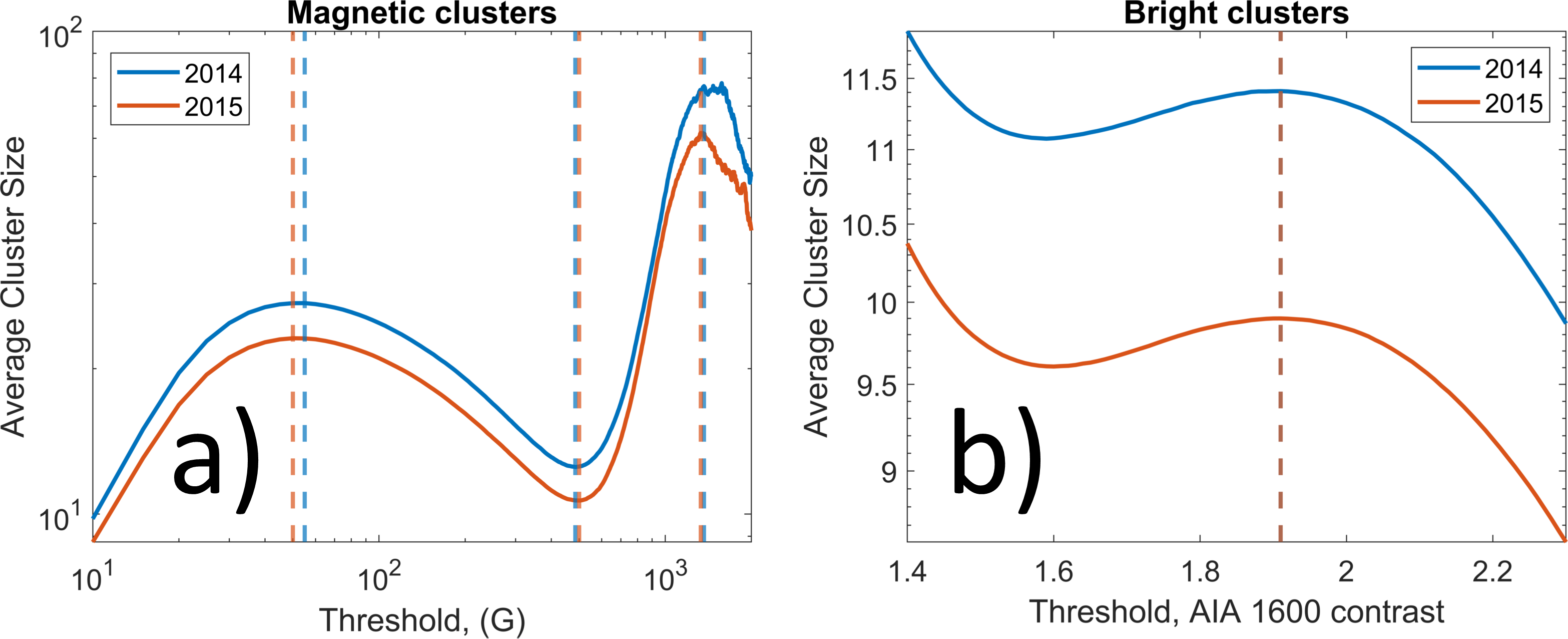}
 \caption{Average size of magnetic (panel a) and bright (panel b) clusters for a given threshold in images from 1 March 2014 to 31 December 2014 (blue) and 1 January 2015 to 31 December 2015 (orange). Vertical lines show locations of maxima and minima.
 For years when the strong field cluster peak is flatter we show the point where the average cluster size flattens.
 \label{fig:ClusterSizes1}}
\end{figure*}

\begin{figure*}
 \centering
 \includegraphics[width=\textwidth,height=\textheight,keepaspectratio]{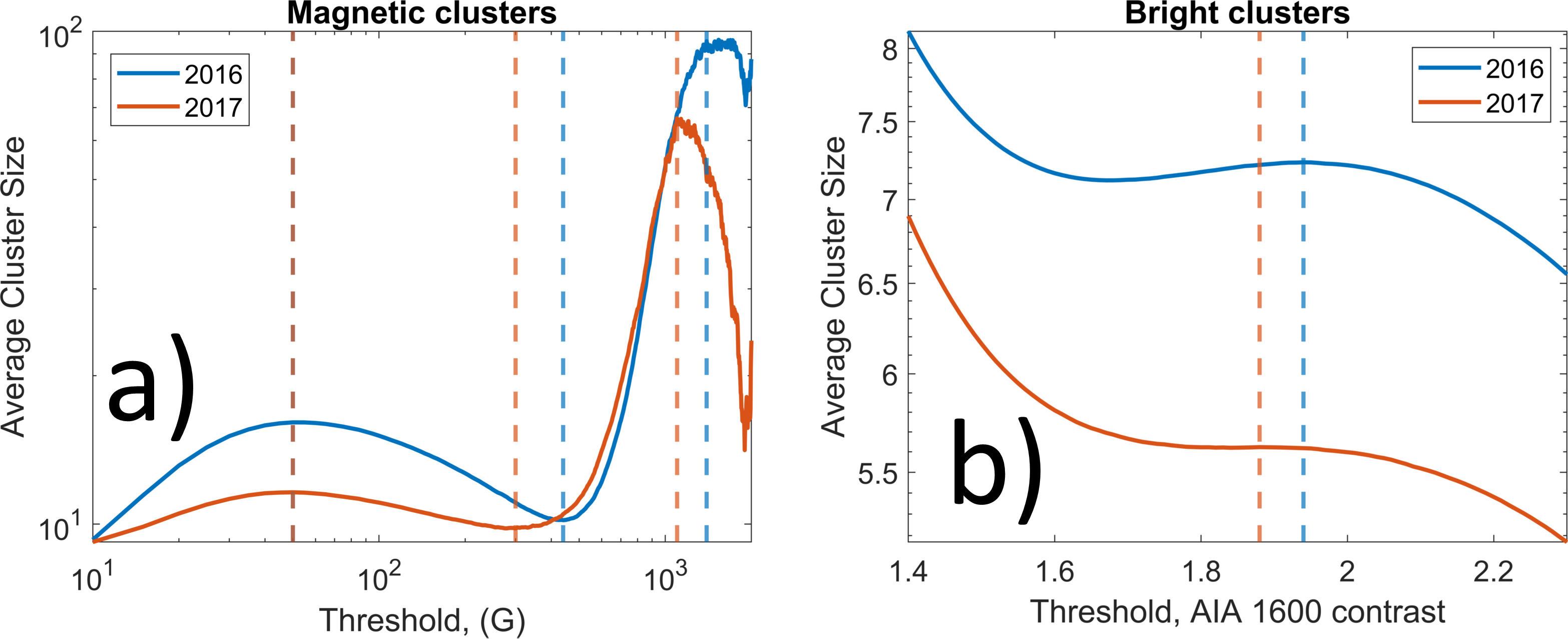}
 \caption{Average size of magnetic (panel a) and bright (panel b) clusters for a given threshold in images from 1 January 2016 to 31 December 2016 (blue) and from 1 January 2017 to 9 June 2017 (orange). The vertical lines show the locations of the maxima and minima.
 For years when the strong field cluster peak is flatter we show the point where the average cluster size flattens.
 \label{fig:ClusterSizes2}}
\end{figure*}

Figure \ref{fig:ClusterDistributions}a shows the average cluster size (in pixels) of magnetic clusters.
We again see that the average magnetic cluster size has a bimodal distribution, this time with the lower peak at about B~=~55~G and the higher peak at about B~=~1365~G, and the intermediate minimum at B~=~475 G.
Similarly, Fig. \ref{fig:ClusterDistributions}b shows the average size of AIA bright clusters as a function of the threshold from I~=~1 to I~=~10.
Again, there is a plateau from around I~=~1.5 to I~=~2, with a small peak at I~=~1.95.
This peak is better seen in Fig. \ref{fig:ClusterDistributions}c, which shows values from I~=~1.5 to I~=~2.2.
Figures \ref{fig:ClusterSizes1} and \ref{fig:ClusterSizes2} show average cluster sizes separately for the years 2014 (from 1 March), 2015, 2016, and 2017 (until 9 June).
The location of the first peak for magnetic clusters stays at 50 -- 55~G, while the average cluster size decreases with decreasing solar activity.
The second peak of magnetic clusters stays in the region from 1330 to 1395~G during first three years and decreases to 1100~G at 2017.
The location of bright cluster peak is in the range 1.88 -- 1.94, although the peak nearly disappears on 2017, when the solar activity is at its lowest.

\begin{figure*}
 \centering
 \includegraphics[width=\textwidth,height=\textheight,keepaspectratio]{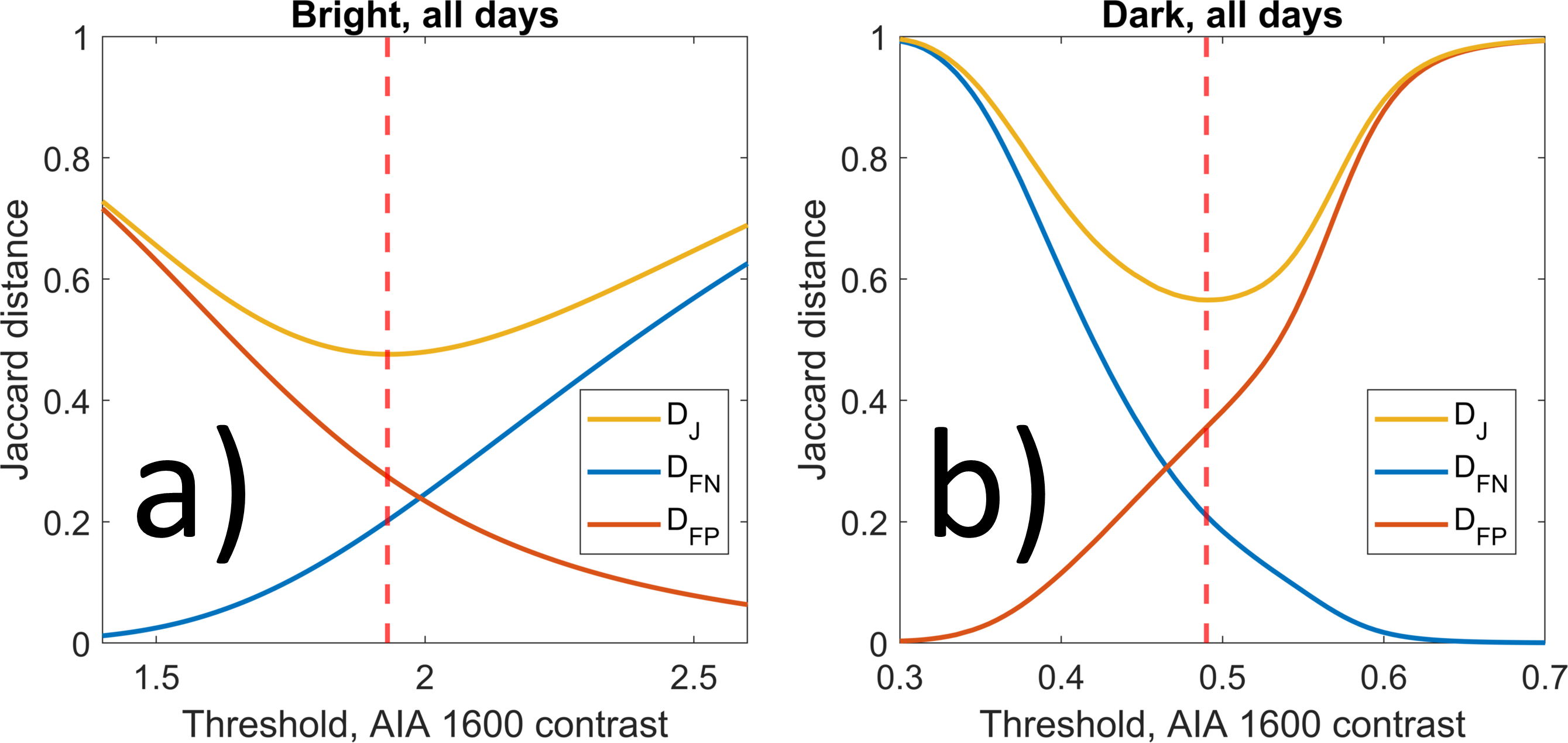}
 \caption{Jaccard distance between AIA and HMI masks as a function of AIA threshold. a) AIA bright pixel masks and HMI moderate field (55 G~$>$~B~$>$~475 G) masks.  b) AIA dark pixel masks and HMI strong field (B~$>$~1365~G) masks. The yellow lines show the value of the  Jaccard distance, while the blue and orange lines show the contributions of  false negatives ($\mathrm{D_{FN}}$) and false positives ($\mathrm{D_{FP}}$).}
 \label{fig:OptimalThresholds}
\end{figure*}

We  show here that the peak at I~=~1.95 corresponds to the peak at B~=~55 G.
With these threshold values, the total areas of bright and magnetic clusters are close to equal (ratio of 0.992).
However, as can be seen from Fig. \ref{fig:ClusterDistributions}, the average size of AIA bright clusters is a factor of two smaller than that of the magnetic clusters, which indicates that the AIA bright clusters are more fragmented than their magnetic counterparts.
To further confirm that we can associate bright and magnetic clusters defined by these thresholds, we studied which AIA threshold would minimize the difference between the masks obtained from bright AIA pixels and from the moderate magnetic field pixels between the lower peak (55 G) and the minimum (475 G).
We excluded the pixels above the minimum from the calculation since intensive magnetic fields typically correspond to decreased radiation.
We compared the masks in the following way:
\begin{enumerate}
    \item For each HMI image we created a mask by setting all pixels with moderate field strength 55~G~$<$~B~$<$~475~G to one and the rest to zero.
    \item For each AIA image we created a set of masks with different contrast thresholds by setting pixels above the threshold to one and the rest to zero.
    \item For each
AIA threshold value we calculated the total Jaccard distance over all days.
\end{enumerate}
The Jaccard distance, also known as the intersection-over-union (IoU) distance, is defined as
\begin{equation}
\begin{split}
    \mathrm{D_{Jaccard}} & = 1 - \mathrm{\frac{AIA \cap HMI}{AIA \cup HMI}~=~1 - \frac{TP}{TP + FP + FN}} \\ 
                         & = \mathrm{\frac{FP + FN}{TP + FP + FN}},
\end{split}
\end{equation}
where $\mathrm{AIA \cap HMI}$ refers to intersection of
AIA and HMI masks and $\mathrm{AIA \cup HMI}$ to their union. The abbreviation
TP stands for true positive, FP for false positive, and FN for false negative pixels, and are  defined such that TP pixels are present in both masks, FP pixels only in AIA masks, and FN pixels only in HMI masks.
The Jaccard distance can be presented as a sum
\begin{equation}
    \mathrm{D_{Jaccard}~=~\frac{FP}{TP + FP + FN}+\frac{FN}{TP + FP + FN}~=~D_{FP}+D_{FN}},
\end{equation}
where $\mathrm{D_{FP}}$ and $\mathrm{D_{FN}}$ measure contributions of false positives and negatives to the Jaccard distance.
Figure \ref{fig:OptimalThresholds}a shows the Jaccard distance (yellow) over all days for bright pixels as a function of the AIA threshold value.
The blue ($\mathrm{D_{FN}}$) and orange ($\mathrm{D_{FP}}$) lines show the contributions of false negatives and positives.
We see that I~=~1.93, which is almost equal to the peak in Fig. \ref{fig:ClusterDistributions}b, minimizes the Jaccard distance between AIA bright pixel masks and HMI moderate field masks.
This result gives compelling evidence that the values around and above the peak I~=~1.95 of Fig. \ref{fig:ClusterDistributions}b indeed correspond to magnetic intensities between the lower peak of Fig. \ref{fig:ClusterDistributions}a and the subsequent minimum.
Figure \ref{fig:OptimalThresholds} also shows how the false positives and negatives contribute to the value of the  Jaccard distance.
With an optimal threshold value their numbers are almost  equal.
We found that I~=~1.93 minimized the Jaccard distance also when we used the location of the second peak, B~=~1365~G, as the upper bound for moderate pixels instead of B~=~475~G.
We repeated the above analysis  for the dark AIA pixels and strong field HMI pixels (B~$>$~1365~G, higher peak of Fig. \ref{fig:ClusterDistributions}a).
Figure \ref{fig:OptimalThresholds}b shows the Jaccard distance with $\mathrm{D_{FP}}$ and $\mathrm{D_{FN}}$ for the dark clusters.
The threshold that minimizes the Jaccard distance between AIA dark pixel masks and HMI strong field masks is I~=~0.49.
This value is close to selected threshold of 0.5, below which there are hardly any AIA pixels on quiet days.

\subsection{Comparison of AIA 1600~\AA{} and HMI masks}
\begin{figure*}
 \centering
 \includegraphics[width=\textwidth,height=\textheight,keepaspectratio]{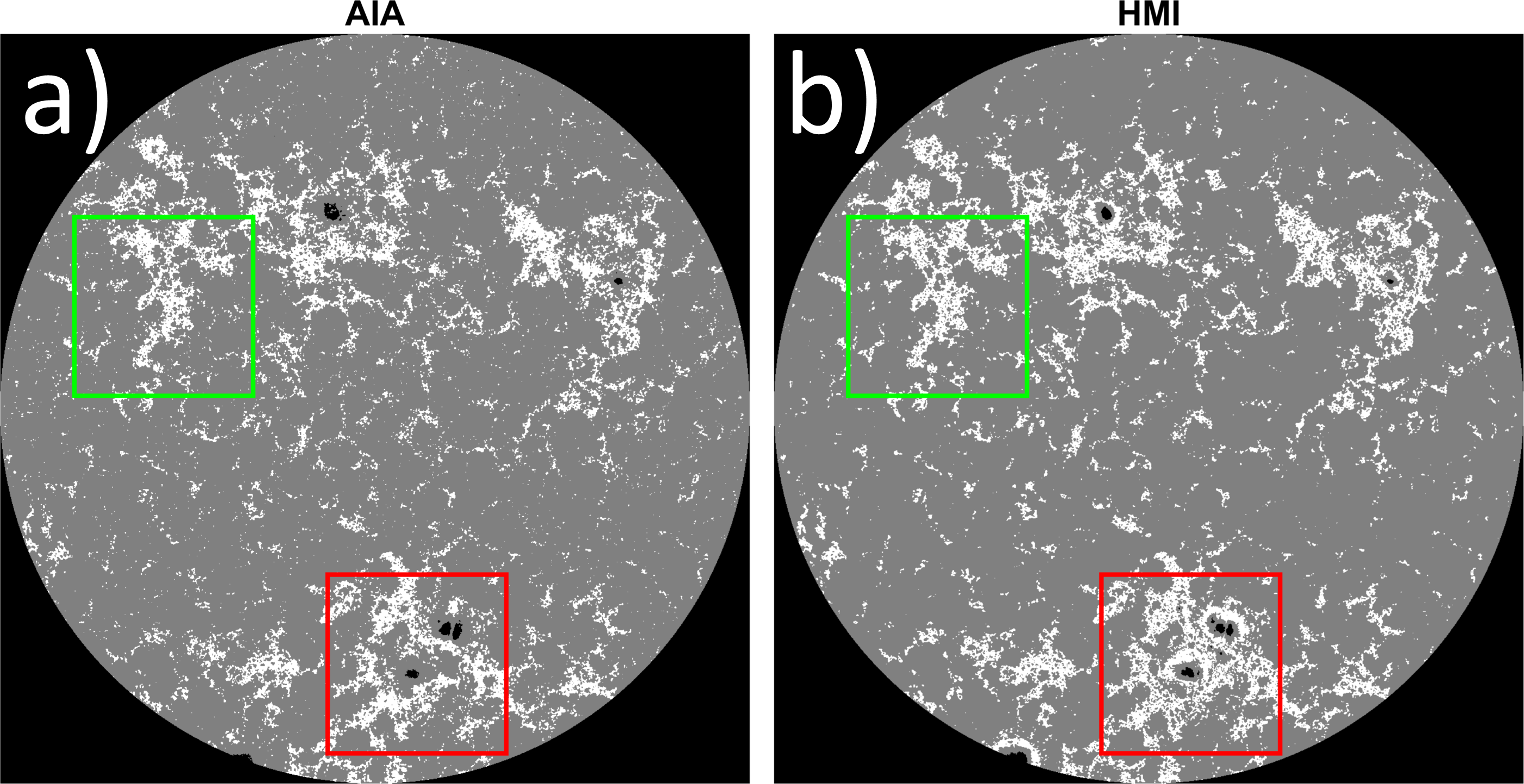}
 \caption{AIA and HMI masks. a) AIA: Bright pixels (I~$>$~1.95) are shown in white and dark pixels (I~$<$~0.5) in black. b) HMI: Moderate field pixels (55 G~$<$~B~$<$~475 G) are shown in white and strong field pixels (B~$>$~1365~G) in black. The shown area corresponds to 0.4R$_\sun$ (see white circle in Fig. \ref{Calibration}, upper panel) for 5 July 2014 (active day). Close-up images of the regions  within the  green and red boxes are shown in Figs. \ref{fig:PlageMasks} and \ref{fig:SunspotMasks}.}
 \label{fig:masks}
\end{figure*}

The minimum Jaccard distance for bright clusters in Fig. \ref{fig:OptimalThresholds}a is 0.4761, while the threshold of I~=~1.95 gives close to an equal value of 0.4763.
For dark clusters the minimum is 0.566, while the threshold I~=~0.5 gives a distance of 0.567.
Figure \ref{fig:masks}a shows an example of bright (I~$>$~1.95) and dark (I~$<$~0.5) pixel masks.
Figure \ref{fig:masks}b shows the  corresponding moderate (55~$<$~B~$<$~475 G) and strong field (B~$>$~1365~G) masks from the HMI image.

\begin{figure*}
 \centering
 \includegraphics[width=\textwidth,height=\textheight,keepaspectratio]{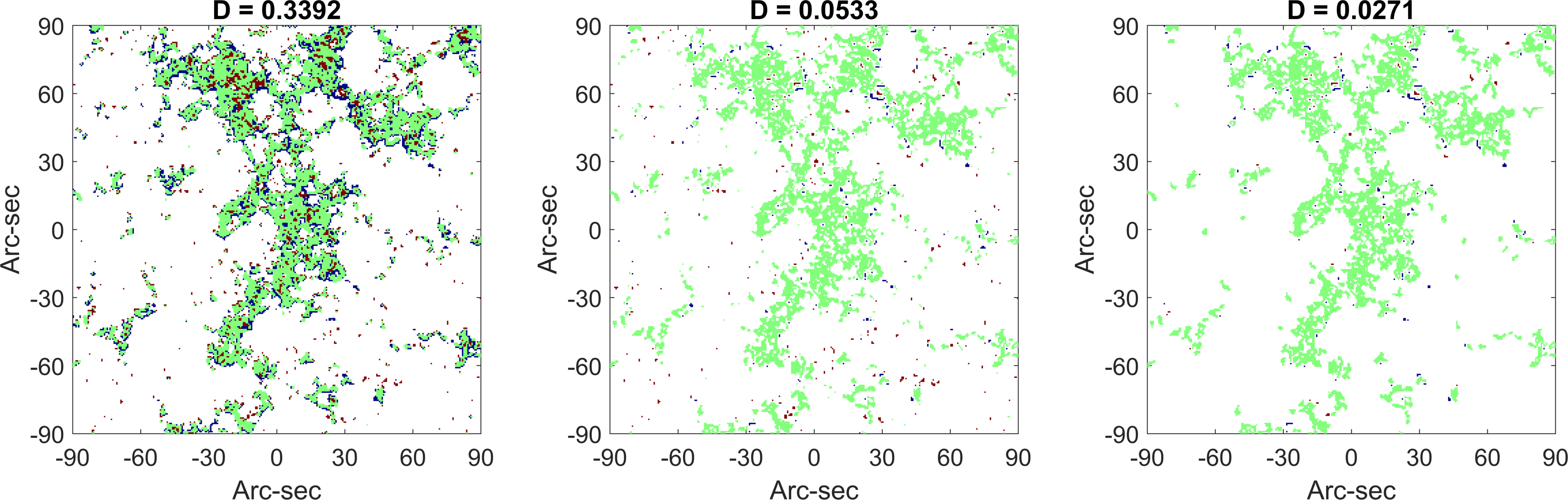}
 \caption{Bright and moderate pixels in plage region. Green: True positive, bright and moderate; Red: False positive, bright not moderate; Blue: False negative, moderate not bright. Left panel: All bright and moderate pixels. Middle panel: Without boundary pixels. Right panel: Without boundary pixels and small clusters (A~<~10~pxl).}
 \label{fig:PlageMasks}
\end{figure*}

Figure \ref{fig:PlageMasks}a shows bright and moderate pixels from the  green box in Fig. \ref{fig:masks}.
This region represents an active region plage without sunspots.
The Jaccard distance between bright and moderate pixel masks within this region is 0.339.
From Fig. \ref{fig:PlageMasks}a we see that much of the bulk of green TP pixels have blue borders indicating that moderate pixel masks tend to extend slightly more than bright pixel masks.
Figure \ref{fig:PlageMasks}a also shows small red regions of FP pixels within otherwise green regions.
These regions correspond to pixels whose magnetic flux density is above B~=~475~G, so they do not appear in moderate field pixel masks.
To examine the effect of border pixels, we excluded FP and FN pixels adjacent to bright and moderate clusters from the calculation of Jaccard distance.
We kept FP and FN pixels more than one pixel away from the cluster boundaries.
Figure \ref{fig:PlageMasks}b shows the masks after the removal of border pixels.
The Jaccard distance decreases  to 0.053, showing that the largest difference between bright and moderate pixel masks arises from the cluster borders.
In the full dataset, 82.8\% of the FN pixels are adjacent to bright clusters, and the total Jaccard distance decreases to 0.234 when we omit the border pixels from the calculation.
Cluster boundaries can have a large effect on the Jaccard distance since both the bright and moderate clusters are fractal-like in the sense that a large fraction of their areas are boundary pixels.

Another discrepancy between bright and moderate pixel masks are small FP bright clusters.
Figure \ref{fig:PlageMasks}c shows the masks after we remove A~<~10~pxl clusters in addition to bordering pixels.
The Jaccard distance further decreases   to 0.027.
The number of FN pixels decreases by a factor of 1.25 and the number of FP pixels by a factor of 5.27.
In the full dataset, 56.9\% of the FP pixels belong to bright clusters smaller than 10 pixels.
The total Jaccard distance decreases to 0.088 when we omit small clusters alongside the boundary pixels from the calculation.

It is important to note that AIA~1600~\AA{} emission and HMI LOS magnetic fields are measured at different altitudes.
AIA 1600~\AA{} emission forms about 400~km above HMI signal (6173~\AA{} line), which is on the order of the  AIA--HMI pixel size (435 km) \citep{Alissandrakis2019}.
When the flux tubes corresponding to the magnetic field measured at a HMI pixel (x,y) bend while arising from the photosphere to the temperature minimum, they might cross the pixel border and appear at the bordering pixel (x$\pm$1,y$\pm$1) in the AIA 1600 \AA{} image.
If the magnetic field on the cluster boundaries tends to incline towards the cluster center, it could explain why the moderate field clusters tend to extend slightly more than the bright clusters.
Furthermore, small AIA 1600~\AA{} clusters, whose field strength is below a moderate level, might be caused by small  subpixel bipolar pairs.
The magnetic fields of such bipolar pairs are averaged out in the  HMI measurements, but they can nevertheless increase AIA 1600~\AA{} emission.

\begin{figure*}
 \centering
 \includegraphics[width=\textwidth,height=\textheight,keepaspectratio]{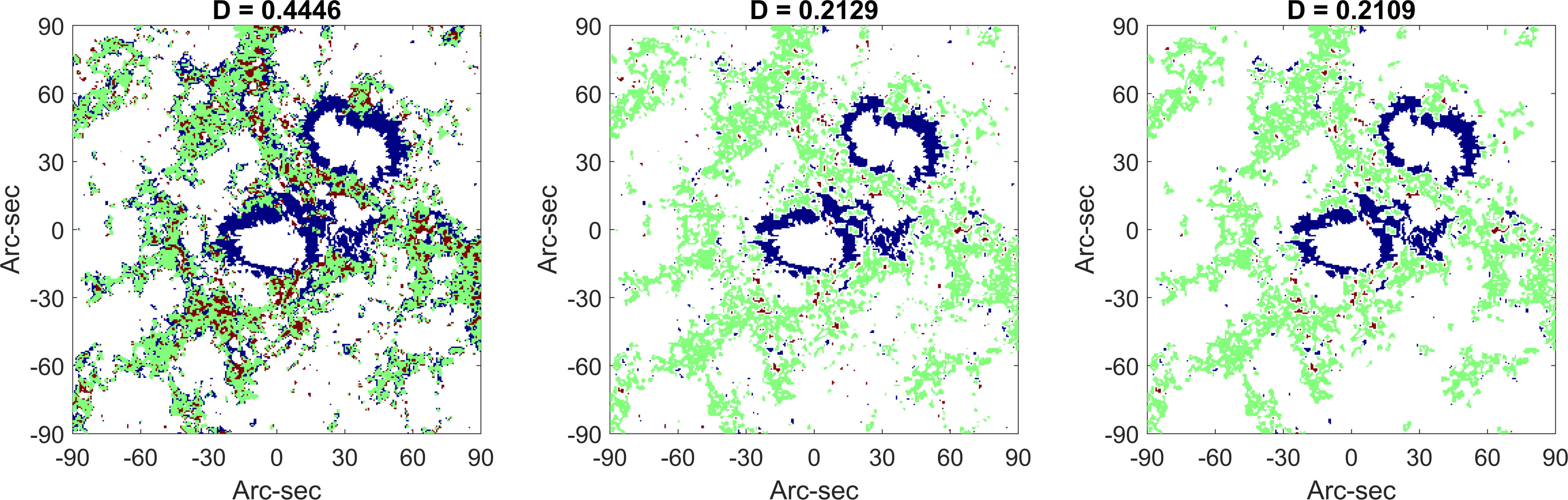}
 \caption{Bright and moderate pixels in sunspot region. Green: True positive, bright and moderate; Red: False positive, bright not moderate; Blue: False negative, moderate not bright. Left panel: All bright and moderate pixels. Middle panel: Without boundary pixels. Right panel: Without boundary pixels and small clusters (A~<~10~pxl).}
 \label{fig:SunspotMasks}
\end{figure*}

\begin{figure*}
 \centering
 \includegraphics[width=\textwidth,height=\textheight,keepaspectratio]{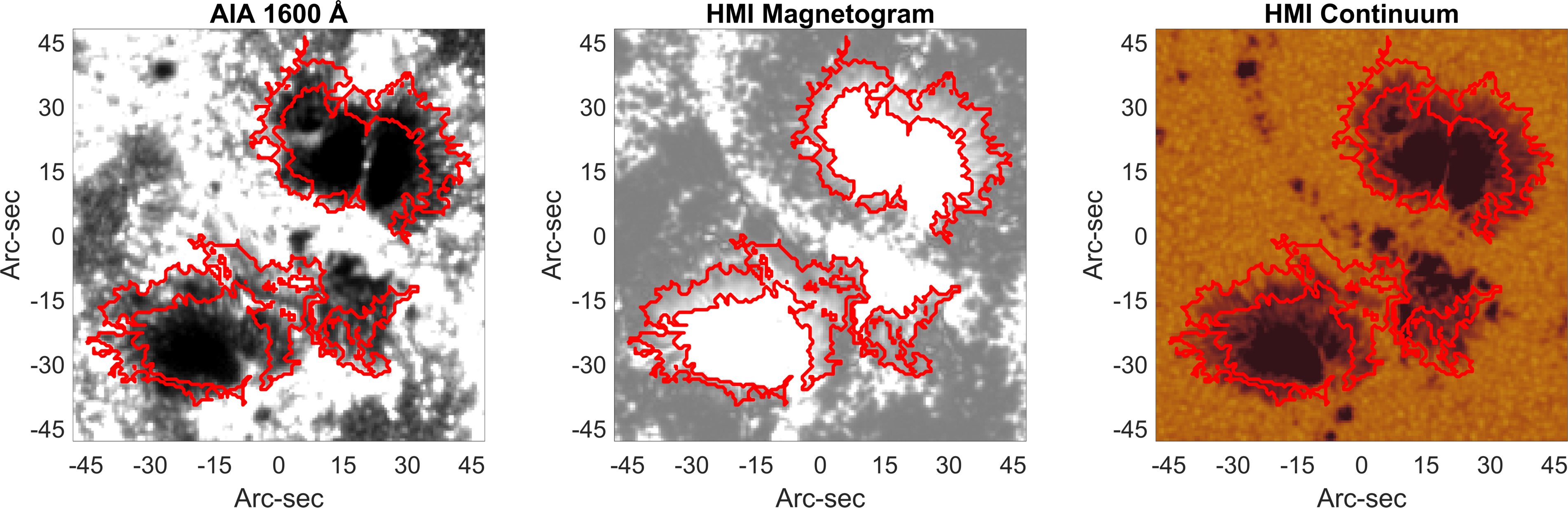}
 \caption{Sunspot region. Left: AIA 1600 \AA{}; middle: HMI magnetogram; right: HMI continuum intensity. The red lines shows boundaries of large 
false negative clusters (A~>~100~pxl). }
 \label{fig:FNBoundaries}
\end{figure*}

\begin{figure*}
 \centering
 \includegraphics[width=\textwidth,height=\textheight,keepaspectratio]{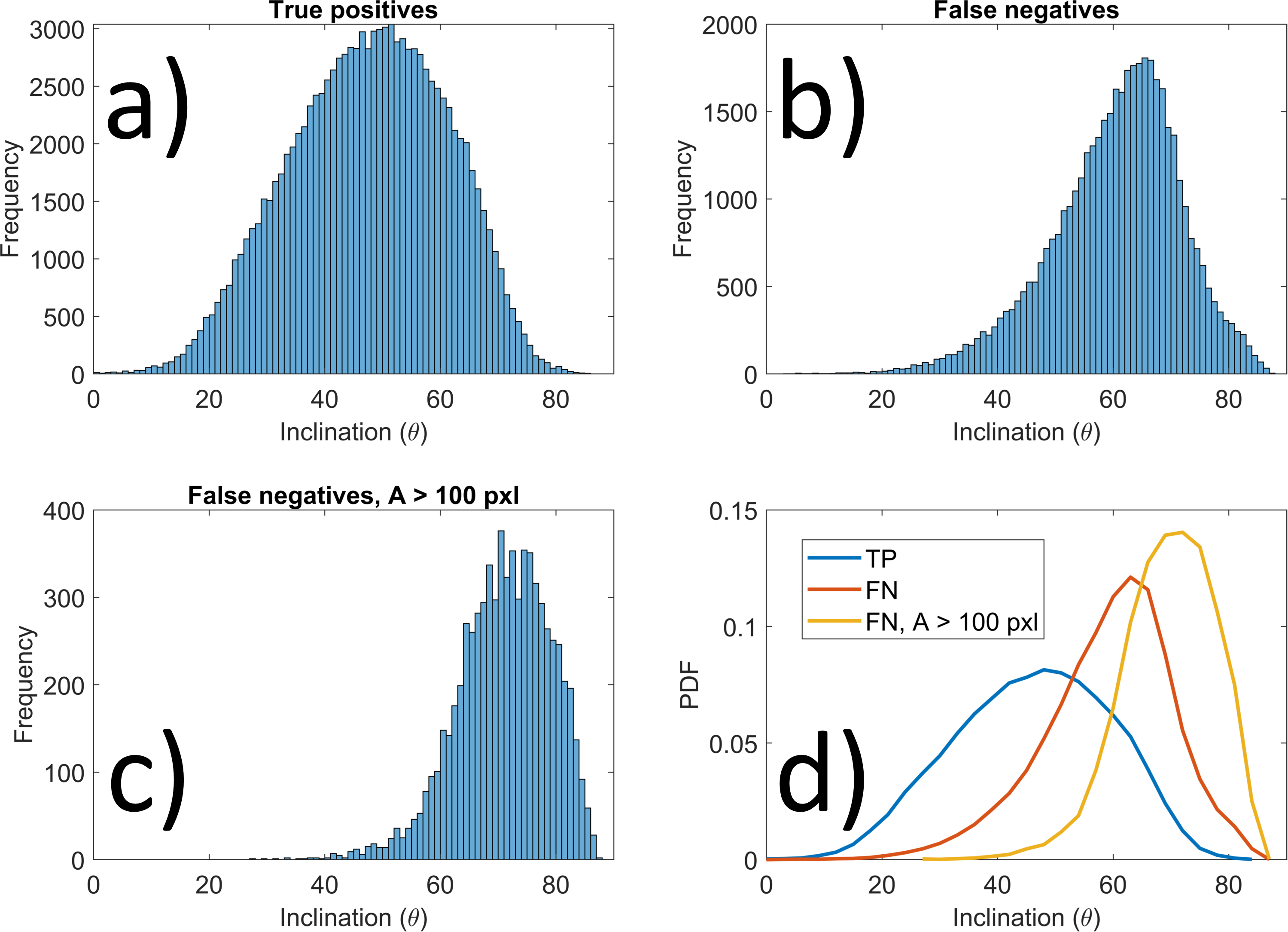}
 \caption{Inclination of magnetic field with respect to the LOS in a) true positive pixels (bright and moderate), b) false negative pixels (moderate not bright),  and c) false negative clusters larger than 100 pixels. Panel d shows the corresponding PDF distributions.}
 \label{fig:InclinationDist}
\end{figure*}

Figure \ref{fig:SunspotMasks} shows an example of a sunspot region, in which we typically see the largest differences between bright and moderate pixel masks.
Panel a shows full bright and moderate pixel masks, panel b shows masks after removing boundary pixels, and panel c after removing boundary pixels and small clusters (A~<~10~pxl).
The Jaccard distance between the masks is 0.445 with all pixels, 0.213 without boundary pixels, and 0.211 without boundary pixels and A~<~10~pxl clusters.
The obvious difference between bright and moderate clusters are large rings of FN pixels around the sunspots.
Figure \ref{fig:FNBoundaries} shows a close-up of the sunspot region in the AIA 1600~\AA{} contrast (panel a), in the  HMI magnetogram (panel b), and in continuum (panel c).
Red lines show boundaries of A~>~100~pxl FN clusters.
These images reveal that large FN clusters correspond to the boundary regions of sunspot penumbrae.
In these regions the strength of the magnetic field belongs to the moderate category, but the AIA pixels do not surpass the bright pixel threshold.
Figure \ref{fig:InclinationDist} shows the inclination of the magnetic field with respect to the LOS for TP and FN pixels for 5 July 2014.
Panel a shows TP pixels, panel b FN pixels, and panel c FN clusters larger than 100 pixels.
Figure \ref{fig:InclinationDist}d shows the corresponding PDFs.
We see that the magnetic field is more horizontal in moderate field pixels, which do not appear bright.
This effect is most pronounced in FN pixels, which form clusters larger than 100 pixels corresponding mainly to sunspot penumbra.
This result also indicates that the small disagreements on the bright and moderate cluster boundaries could be caused by the more horizontal magnetic field.
We note that the analysis of the magnetic field inclination was based on a single day, 5 July 2014.

\begin{figure*}
 \centering
 \includegraphics[width=\textwidth,height=\textheight,keepaspectratio]{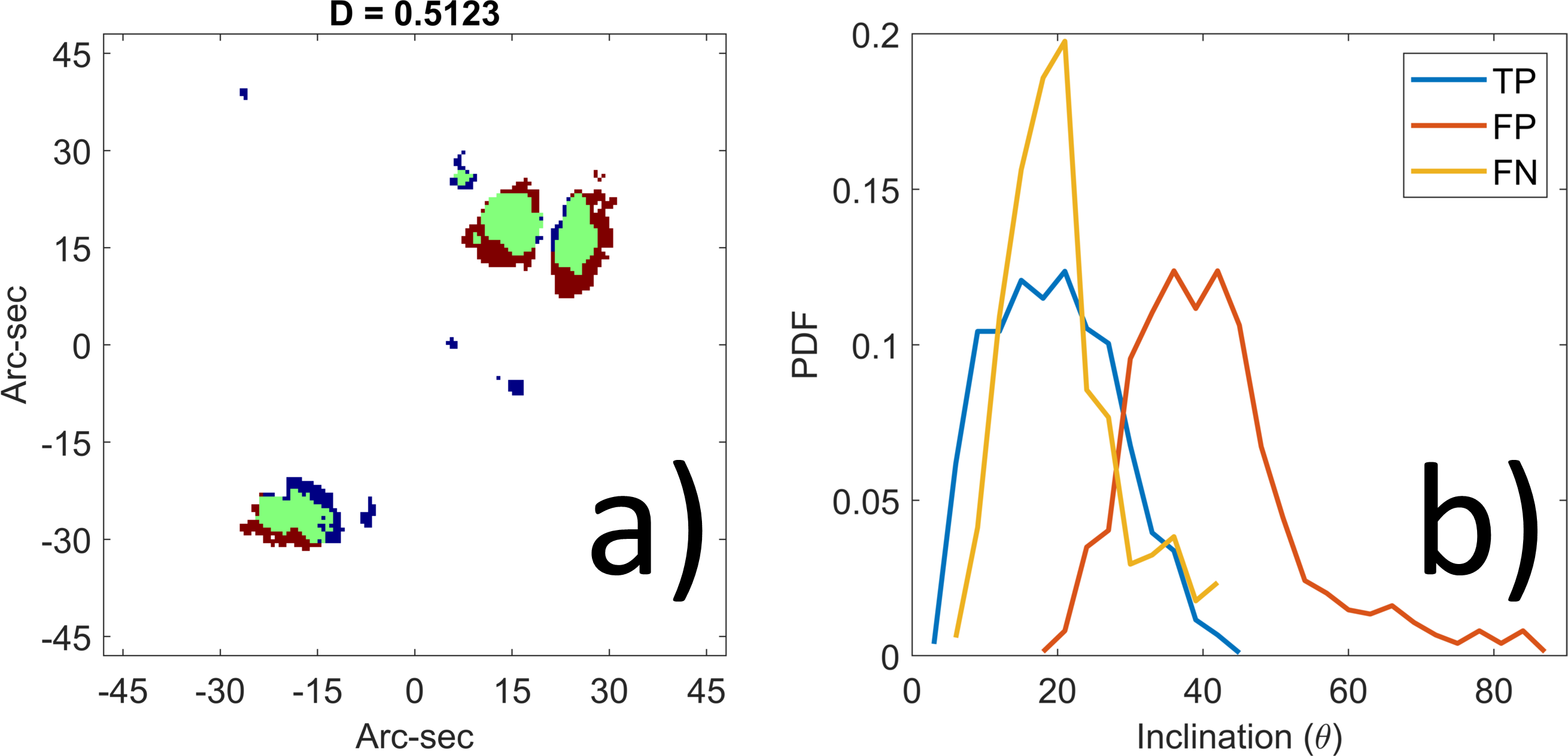}
 \caption{Dark and strong field pixels. a) Close-up of sunspot region. Green: True positive, dark and strong; Red: false positive, dark not strong; Blue: False negative, strong not dark. b) Inclination of magnetic field with respect to the LOS in true positive pixels (blue), false positive pixels (orange), and false negative pixels (yellow).}
 \label{fig:DarkStrongMasks}
\end{figure*}

The nature of dark and strong field clusters is somewhat different from bright and moderate clusters.
They are more uniform (i.e., less fractal-like or fragmented), which is also reflected by the  larger average size of strong field clusters in Figs. \ref{fig:ClusterDistributions} -- \ref{fig:ClusterSizes2}.
Figure \ref{fig:DarkStrongMasks}a shows masks of dark and strong field pixels from the sunspot region discussed above.
There is a disagreement between cluster boundaries, including both FN and FP pixels.
Figure \ref{fig:DarkStrongMasks}b shows PDF distributions for magnetic field inclination for TP, FN, and FP pixels.
The distribution of FP pixels (dark, not strong)  clearly differs from those of TP and FN pixels.
The magnetic field is close to radial in TP and FN pixels, but it is more horizontal in FP pixels.
Similarly, as with bright and moderate pixels, this indicates that a more horizontal magnetic field decreases AIA 1600 \AA{} radiation.

\begin{figure*}
 \centering
 \includegraphics[width=\textwidth,height=\textheight,keepaspectratio]{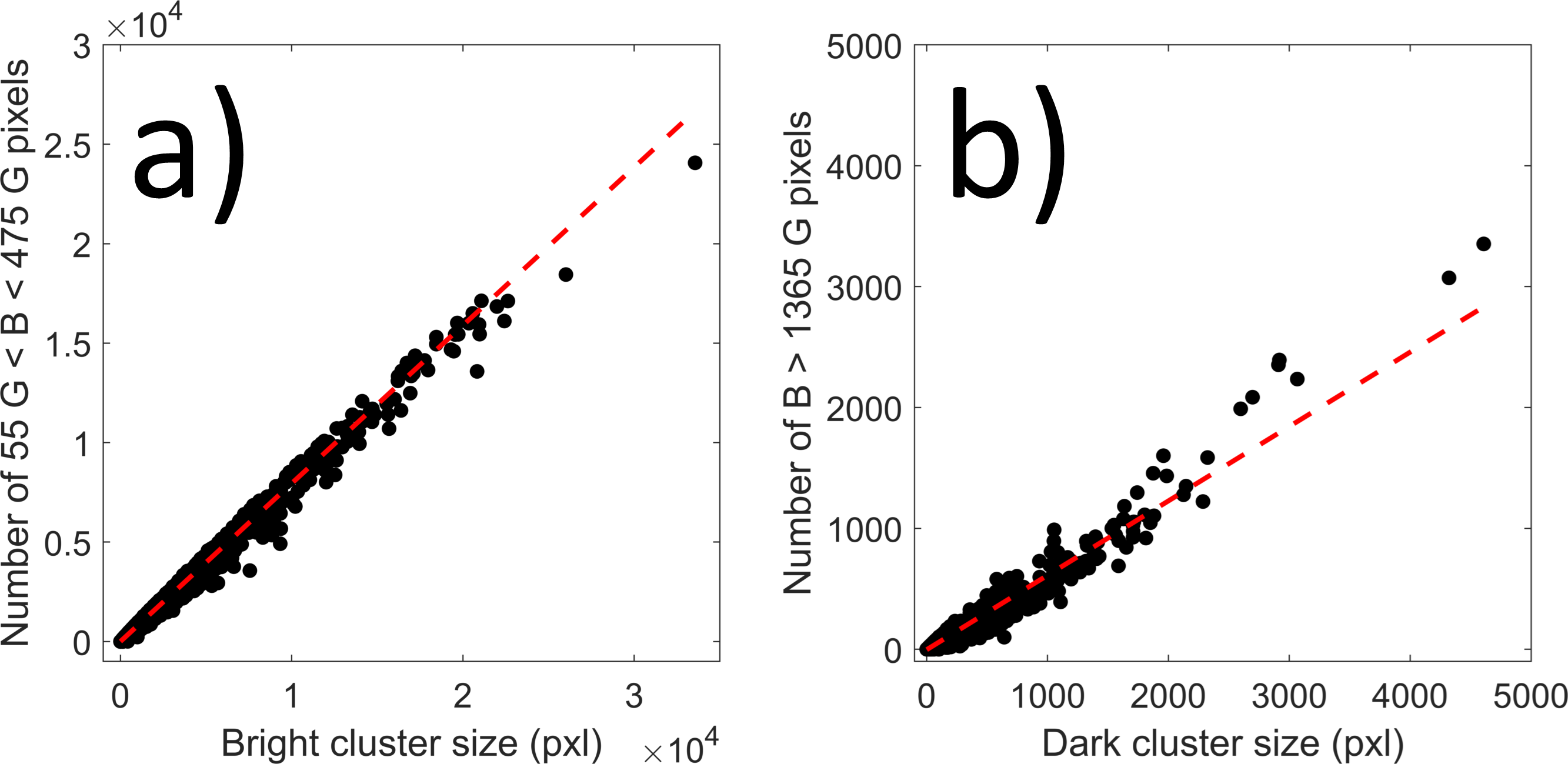}
 \caption{a) Number of moderate field pixels (55 G$<$ B~$<$~475 G)  within bright clusters as a function of cluster size. b) Number of strong field pixels (B~$>$~1365 G)  within dark clusters as a function of cluster size. The red lines show the least-squares linear fits.}
 \label{fig:NumberOfMagneticPixels}
\end{figure*}

Figure \ref{fig:NumberOfMagneticPixels}a shows the number of moderate magnetic field pixels (55 G~$<$~B~$<$~475 G)  within the bright cluster as a function of the cluster size.
The red line shows the least-squares linear fit ($R^2~=~0.993$).
The number of magnetic pixels within a bright cluster is linearly proportional to the cluster size with a slope of 0.795.
Thus, 79.5 \% of the bright clusters consist of moderate magnetic field pixels.
Table \ref{table:MagneticFractionsBright} shows fractions of magnetic pixels within bright clusters with different upper and lower bounds.
The reported fractions are based on the slope of the regression line.
For B~>~1365 pixels fitting is not feasible ($R^2~=~0.01$) due to their extremely small number within bright clusters (0.002\% in total).
From Table \ref{table:MagneticFractionsBright} we see that 12.9 \% ($R^2~=~0.913$) bright cluster pixels are associated with field strengths above 475 G.
If we take these into account, we find that the fraction of magnetic pixels within bright clusters rises to 92.4\% ($R^2~=~0.996$).
Figure \ref{fig:NumberOfMagneticPixels}b shows the number of strong field pixels (B~$>$~1365 G)  within the dark cluster as a function of the cluster size.
The red line shows the least-squares linear fit ($R^2~=~0.953$).
The slope of the regression line is 0.615, so 61.5 \% of dark clusters consist of strong magnetic field pixels.
Table \ref{table:MagneticFractionsDark} shows the fraction of magnetic pixels within dark clusters with different upper and lower bounds.
If we also  take into account  pixels above B~=~475~G (the minimum of \ref{fig:ClusterDistributions}a), the fraction increases to 98.8\% ($R^2~=~0.9996$).
While moderate field pixels are almost exclusively bright and strong, field pixels are almost exclusively dark; the intermediate range from B~=~475~G to B~=~1365 G represents a mixed region containing both bright and dark pixels.

\begin{table*}
\caption{Fraction of pixels within given magnetic range in bright clusters.} 
\label{table:MagneticFractionsBright} 
\centering 
\begin{tabular}{c|c c c} 
\hline\hline 
\backslashbox{Lower bound (G)}{Upper bound (G)} & 475 & 1365 & $\infty$ \\ 
\hline 
        
55      & 0.7949 & 0.9241 &  0.9243 \\
475     &    -   & 0.1293 &  0.1295 \\

\hline
\end{tabular}
\end{table*}

\begin{table*}
\caption{Fraction of pixels within given magnetic range in dark clusters.} 
\label{table:MagneticFractionsDark} 
\centering 

\begin{tabular}{c|c c c} 
\hline\hline 
\backslashbox{Lower bound (G)}{Upper bound (G)} & 475 & 1365 & $\infty$ \\ 
\hline 
55      & 0.0115 & 0.3840 &  0.9996 \\
475     &    -   & 0.3726 &  0.9881 \\
1365   &    -   &    -   &  0.6155 \\

\hline
\end{tabular}
\end{table*}

\subsection{Final thresholds}
Based on our results, we set  $I_{\mathrm{BP}}~=~1.95$ as the threshold to define the bright pixels and $I_{\mathrm{BP}}~=~0.5$ as the threshold for dark pixels.
These threshold values were obtained solely from the AIA data, but they are very close to the values obtained from the comparison of the  AIA and HMI masks (I~=~1.93 and B~=~0.49; Fig. \ref{fig:OptimalThresholds}).
These results also show that we can associate bright and dark clusters with moderate (B~$>$~55 G) and strong  (B~$>$~1365~G) magnetic clusters, respectively.
The two peaks in Fig. \ref{fig:ClusterDistributions}a correspond to true magnetic thresholds, which maximize the average size of magnetic clusters of the 4-connected  HMI pixels exceeding the given threshold.
The bimodal structure of the average cluster sizes can be understood as follows.
Although small threshold values (B~<~55 G) produce the largest magnetic clusters, they also lead to a large number of small and scattered clusters which reduces the average size.
From 55~G up to 475~G, the average cluster size decreases since the large clusters fragment and shrink faster than the number of small clusters decreases.
Above 475~G the average cluster size increases, indicating that small clusters begin to disappear rapidly.
The threshold of 1365~G corresponds to the field strength of pores and sunspot umbrae, which have fairly narrow area distribution (see later in Sect. \ref{sec:clusters}).
The peak at 55~G is in agreement with the results of \citet{Schrijver1987} who, using the 40-channel magnetograph \citep{Livingston1971} at Kitt Peak National Observatory, found that the magnetic threshold of 50~G produced the most coherent set of clusters (magnetic plages).
Smaller thresholds led to a large set of many small clusters while higher thresholds led to more fragmented clusters.

As already indicated by their smaller average size, AIA 1600~\AA{} bright clusters are more fragmented than their magnetic counterparts.
The most evident difference between these masks can be seen around dark and strong field clusters (in  black in both images).
In HMI masks, strong field pixels are surrounded by moderate field pixels, seen as white rings around black strong field pixels.
These surrounding pixels correspond to sunspot penumbrae, which mostly do not appear as bright pixels in the AIA mask.
In comparison with solar features, it appears that moderate field (bright) pixels within a range of  55--475~G (I $\geq$ 1.95)  typically correspond to solar plages and the enhanced network and they also display the supergranulation pattern.
Strong field (dark) pixels with B$~>~$1365~G (I~<~0.5) correspond to sunspot umbrae and pores.
Pixels within 475--1365~G correspond to sunspot penumbrae, and pixels with the weakest magnetic field (B$<$55~G) form the network flux.

\begin{figure*}
 \centering
 \includegraphics[width=\textwidth,height=\textheight,keepaspectratio]{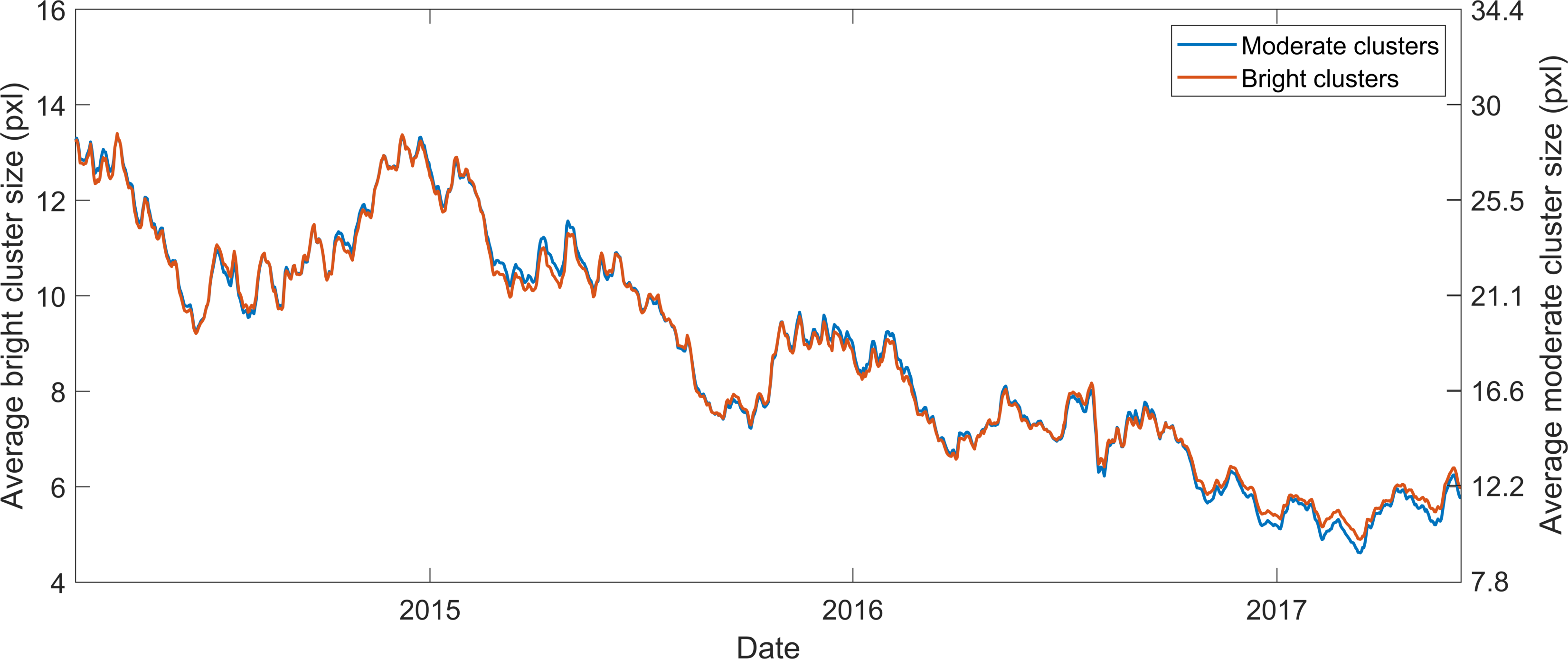}
 \caption{Average cluster sizes of bright clusters (orange) and moderate clusters (blue). Daily values are smoothed with a 27-day moving window.
 }
 \label{fig:AverageClusterSize}
\end{figure*}

In previous studies, many different thresholds were used to separate bright features from the rest of the solar disk.
Often, these thresholds are dynamical,  functions of individual image properties such as the mean and standard deviation of intensity of the image.
However, how  such a threshold function is defined is  somewhat subjective, although  more objective methods have been proposed \citep{Ermolli2007}.
\citet{Bose2018}, who also studied AIA 1600~\AA{} data, relied on a dynamical threshold that changes from image to image and questioned the use of a constant threshold altogether.
They argue that a constant threshold gives inconsistent results over a longer period because of the time-varying image properties.
While this argument may be particularly true for ground data and especially for historical spectroheliograms \citep{Ermolli2009,Chatzistergos2019a}, we argue that it does not hold for high-quality observations made by SDO.
We provide evidence for our contrary view in Fig. \ref{fig:AverageClusterSize}, which shows the smoothed (27-day moving average) time series of the average cluster size of bright AIA clusters and moderate clusters (55~$<$~B~$<$~475 G).
The sizes are given in pixels and shown on the  left for bright clusters and on the  right for moderate clusters.
Figure \ref{fig:AverageClusterSize} shows that the variation of the average bright cluster size is almost indistinguishable from that of the moderate cluster size throughout the studied period.
The two cluster sizes depict, in a very similar way, considerable short-term variability and an obviously solar cycle related decrease where the average size is related roughly by a factor of two.
This excellent agreement between two completely independent datasets shows that the adopted calibration method corrects the inter-image variability, and no further measures, such as temporally variable (dynamical) thresholding are needed.
The selected contrast thresholds produce the most coherent bright and dark clusters and they correspond to magnetic thresholds that maximize the average size of moderate and strong field clusters in HMI magnetograms.

\section{Disk-integrated AIA 1600~\AA{} contrast and HMI magnetic field}
\label{sec:disc_int}
\subsection{Effect of decreased HMI noise level}
\begin{figure}
\centering
\includegraphics[width=\columnwidth,height=\textheight,keepaspectratio]{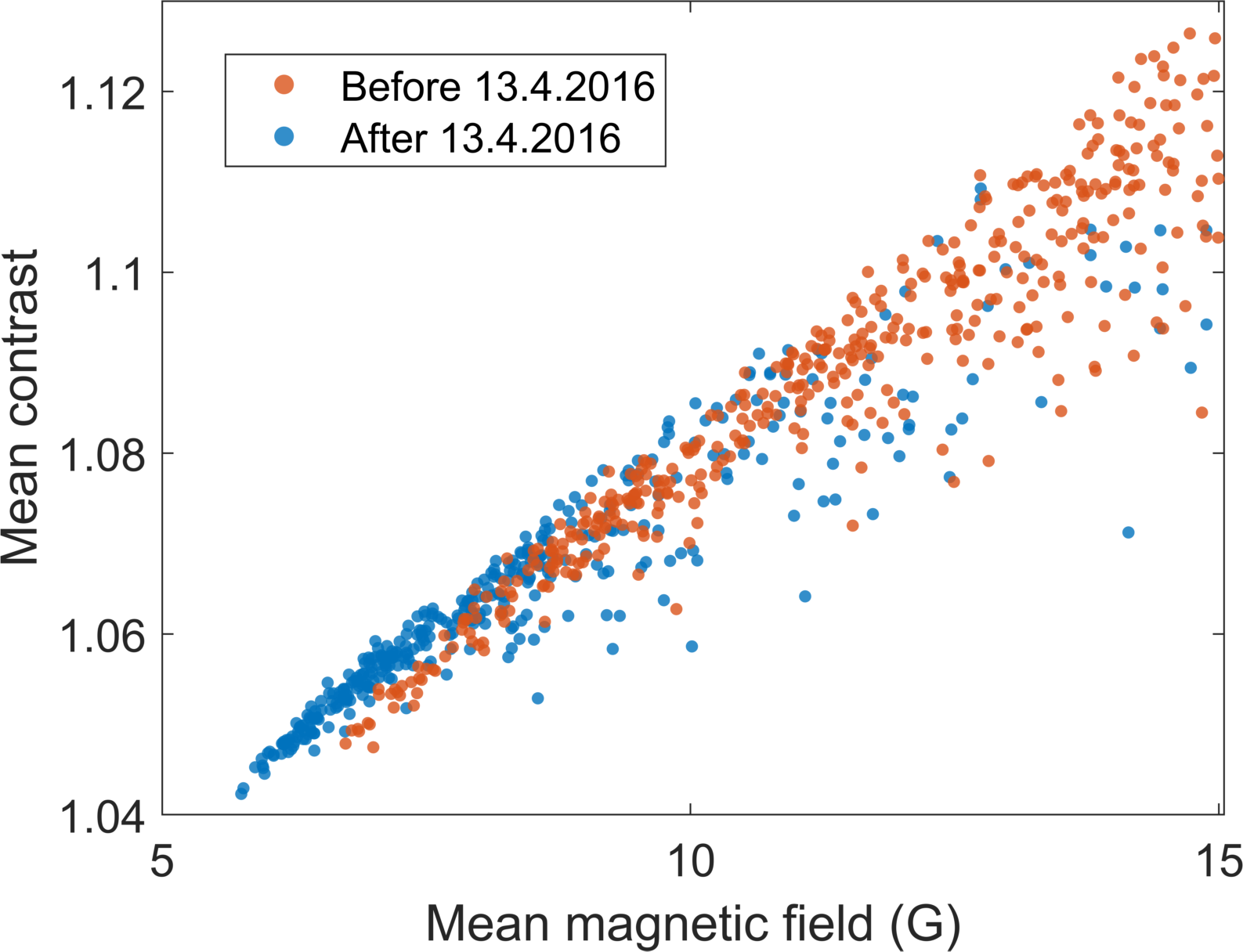}    
\caption{Effect of decreased HMI noise level. Mean AIA 1600~\AA{} calibrated contrast as a function of mean unsigned magnetic field within $0.4R_{\sun}$. The days before 13 April 2016 are shown in blue, and the days after  13 April 2016 in orange. The axes have been cut short to bring out the difference of the small values.
\label{fig:HMINoise}}%
\end{figure}

The HMI changed the scheme for collecting vector magnetic field measurements on 13 April 2016 \citep[][see also HMI Science Nuggets \#56\footnote{\url{ http://hmi.stanford.edu/hminuggets/?p=1596}}]{Liu.etal2016}.
This change decreases the HMI noise level and affects the measurements of weak magnetic fields.
We can see the effect of this change in Fig. \ref{fig:HMINoise}, which shows the scatter plot of the mean calibrated 1600~\AA{} contrast as a function of the mean unsigned magnetic field within the 0.4 $R_{\sun}$ disk (which we  use here, as noted earlier).
After 13 April 2016 the average values, especially those below   10~G, are shifted towards weaker magnetic field values, due to the decreased noise level.
Since our  aim was to study the relation between AIA 1600~\AA{} contrast and magnetic fields in fairly active times, and the change in the HMI measurement scheme only affects the weakest fields, we mostly ignored this change in the HMI measurements.
However, since this change affects the disk mean values, in this section we treat the two periods  separately.
The pre-13 April 2016 period has 765 images and the post-13 April 2016 has 400 images.

\subsection{All pixels}
  \begin{figure*}
   \centering
  \includegraphics[width=\textwidth,height=\textheight,keepaspectratio]{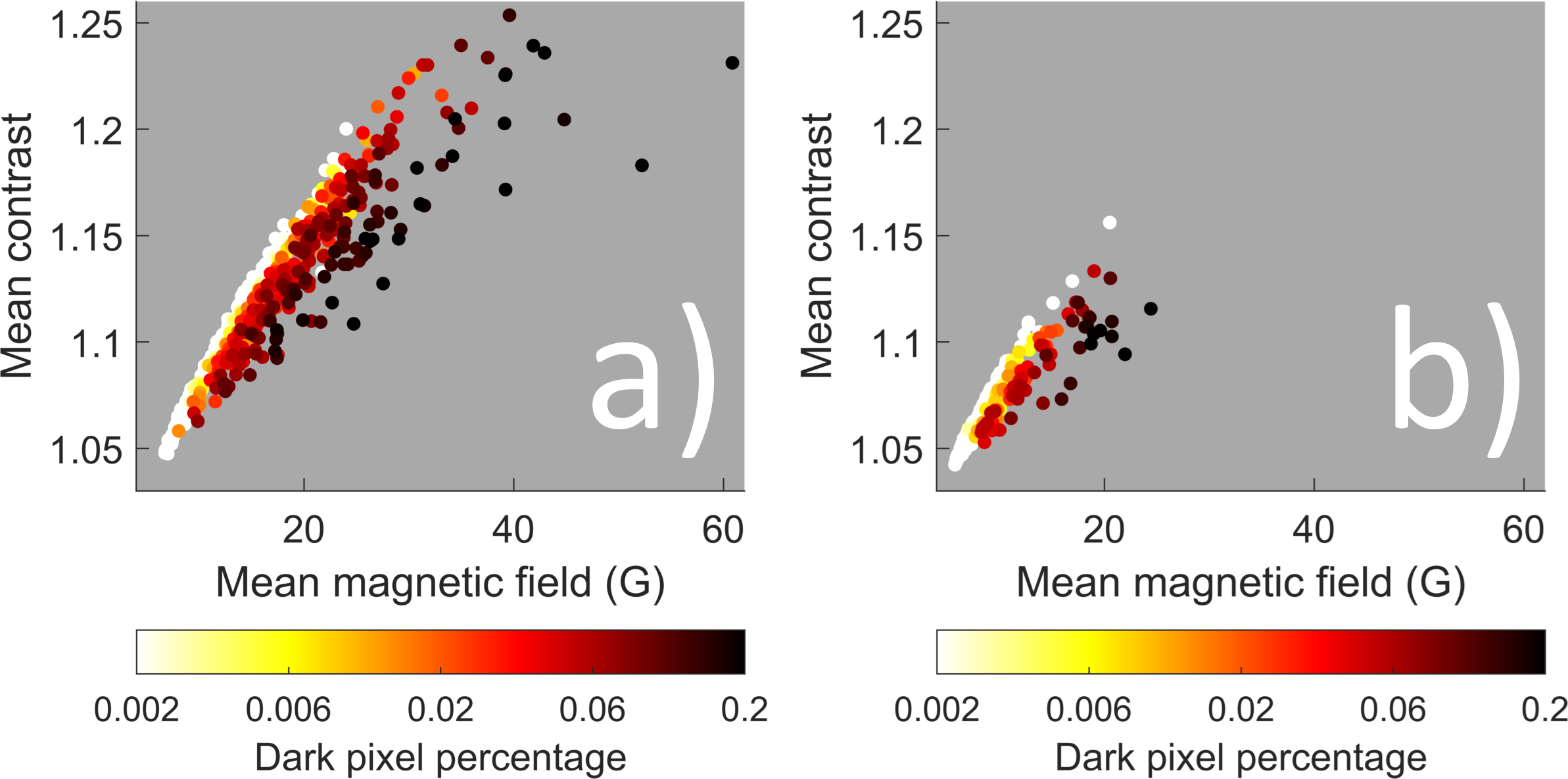}
   \caption{Mean AIA 1600~\AA{} calibrated contrast as a function of mean unsigned magnetic field a) before and b) after HMI measurement scheme change. The color of the  points indicates the percentage of solar disk covered by the dark pixels. \label{MeanAIAvsHMI}}%
    \end{figure*}

Figure \ref{MeanAIAvsHMI} shows the scatter plots for the mean 1600~\AA{} contrast and the mean unsigned magnetic field using all pixels within a $0.4R_{\sun}$ radius from the disk center before  (panel a) and after (panel b) the HMI measurement scheme change.
We  color the points by the percentage of dark pixels appearing in the image.
Figure \ref{MeanAIAvsHMI} shows that the mean AIA 1600~\AA{} contrast mainly increases with the magnetic field, but there is also some variability.
For example, most of the days in Fig. \ref{MeanAIAvsHMI}a with the mean AIA 1600~\AA{} contrast between 1.15  and  1.2 have a mean magnetic field in the range 20--30~G, but there is also one day for which B~=~52 G.
From the color of the points, we also see that the percentage of dark pixels increases with the mean magnetic field, as indicated by the dark color of the most magnetically intensive days.
Accordingly, the variability in the mean contrast--mean unsigned magnetic field relationship primarily comes from the percentage of dark pixels visible on the solar disk.
For a constant mean magnetic field, a smaller dark pixel percentage indicates a higher mean contrast.
Conversely, for a constant mean AIA 1600~\AA{} contrast, a higher dark pixel percentage indicates a higher mean magnetic field.
Figure \ref{MeanAIAvsHMI}a shows that the range of mean contrast and magnetic field values is much higher during the period before the HMI measurement scheme change than during the later period.
The earlier period covers the maximum and declining phases of the solar cycle, while the latter covers only the declining phase.
    
      \begin{figure*}[htbp!]
   \centering
  \includegraphics[width=\textwidth,height=\textheight,keepaspectratio]{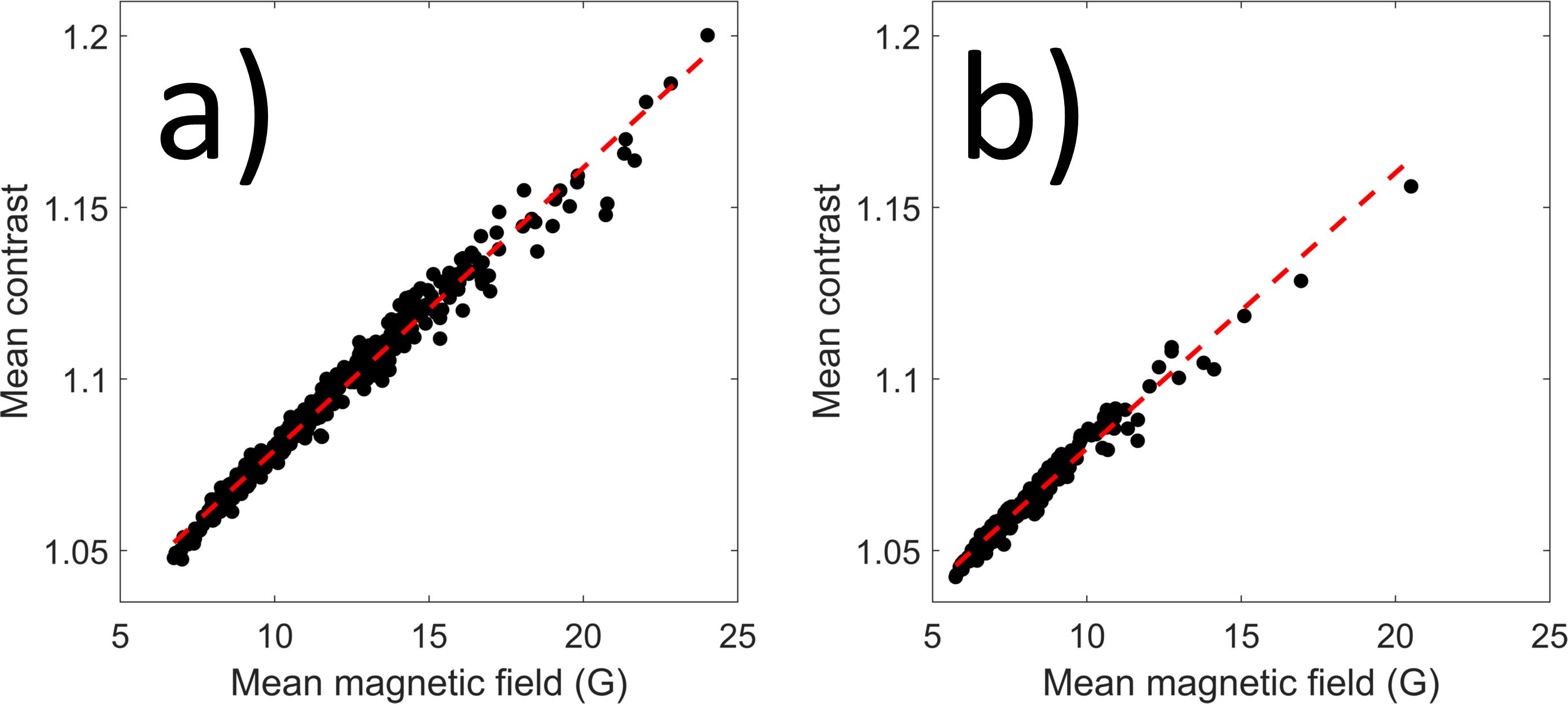}
   \caption{Mean AIA 1600~\AA{} calibrated contrast as a function of mean unsigned magnetic field a) before and b) after the  HMI measurement scheme change on 13 April 2016. Only days with a  dark pixel percentage less than 0.001\% are shown. The red lines show the best least-squares regression lines. The best-fit parameters are shown in Table \ref{table:I_vs_B}. \label{MeanAIAvsHMIFit}}%
    \end{figure*}

\begin{table*}[hbt!]
\caption{Regression coefficients of the model $\langle I \rangle~=~\beta_{0} +  \beta_{1}\langle B \rangle$.} 
\label{table:I_vs_B} 
\centering 

\begin{tabular}{c|c c|c c} 
\hline\hline 
Time & $\beta_0$ & $\beta_1$ & $R^2$ & MAD \\ 
\hline 
    
Before  &   0.9968$\pm$0.0008 &   0.00806$\pm$0.00006   &   0.980   &   0.0029     \\
After   &   0.9995$\pm$0.0006 &   0.00803$\pm$0.00008   &   0.973   &   0.0017     \\

\hline
\end{tabular}
\end{table*}

Figure \ref{MeanAIAvsHMIFit} shows the scatter plots from Fig. \ref{MeanAIAvsHMI} for only those days for which the dark pixel percentage is less than 0.001\%, together with the best-fit least-squares line.
Most points fall close to the regression line, indicating that when the dark pixel percentage is close to negligible, the relationship between the mean contrast and the mean unsigned magnetic field is close to linear.
A small amount of  nonlinearity seems to be present at the weakest magnetic field values, especially before the HMI scheme change (Fig. \ref{MeanAIAvsHMIFit}a).
We show the best-fit parameters in Table \ref{table:I_vs_B}.
The intercept parameters of these fits are almost equal and are close to unity.
The regression line slope is somewhat larger in the later period than in the earlier period.
The mean absolute deviation (MAD) from the regression line is 0.0029 for the earlier period and 0.0017,  almost two times less, for the later period.
These differences are due to the change in the HMI measurement scheme.
The decreased noise level after the  HMI change shifts magnetic intensities to smaller values so that the effect is greater on quiet than active days, making the fit more linear.
Accordingly, our results support the adopted HMI scheme change.
We note that $R^2$ is slightly higher for the earlier period due to a larger range of contrast values.

\subsection{Bright and dark pixel percentages}
\begin{figure*}[hbt!]
   \centering
  \includegraphics[width=\textwidth,height=\textheight,keepaspectratio]{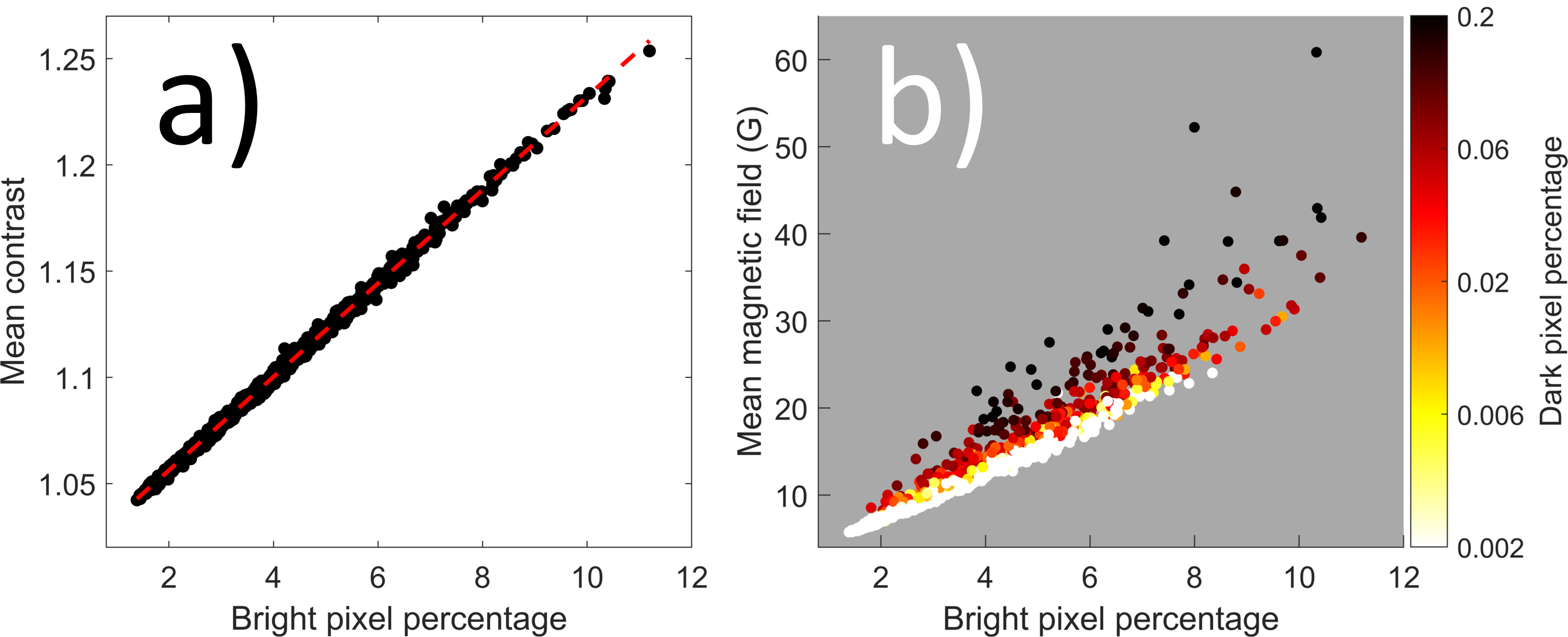}
   \caption{Bright pixel relations. a) Mean AIA 1600~\AA{} calibrated contrast as a function of bright pixel percentage. The red line shows the best least-squares regression line. The best-fit parameters are shown in Table \ref{table:I_vs_Bp}. b) Bright pixel percentage as a function of the mean unsigned magnetic field. The color of the points indicates the percentage of solar disk covered by the dark pixels (I~<~0.5). \label{BrightPixels}}%
    \end{figure*}

Figure \ref{BrightPixels}a shows the scatter plot of the mean AIA 1600~\AA{} contrast as a function of bright pixel percentage.
The best-fit parameters are shown in Table \ref{table:I_vs_Bp}.
We see that there is a strongly linear relation over the whole contrast range.
This relation shows that the fraction of bright pixels on the solar surface almost entirely explains the variability in the mean AIA 1600~\AA{} contrast.
Notably, the days when the mean AIA 1600~\AA{} contrast saturates in Fig. \ref{MeanAIAvsHMI} (i.e., those that deviate the most from the linear relation), do not stand out in Fig. \ref{BrightPixels}a.
This lack of saturation in the bright pixel percentage--mean contrast relation indicates that a small mean contrast during some days is due to the small percentage of bright pixels, not due to an excess of dark pixels.

\begin{table*}[hbt!]
\caption{Regression coefficients of the model $\langle I \rangle~=~\beta_{0} + \beta_{1}f_{BP}$.} 
\label{table:I_vs_Bp} 
\centering
\begin{tabular}{c c|c c} 
\hline\hline 
$\beta_0$ & $\beta_1$ & $R^2$ & MAD \\ 
\hline 

1.0123$\pm$0.0001 &   0.02199$\pm$0.00003       &   0.998   &   0.0013 \\
\hline
\end{tabular}
\end{table*}

\begin{figure*}
   \centering
  \includegraphics[width=\textwidth,height=\textheight,keepaspectratio]{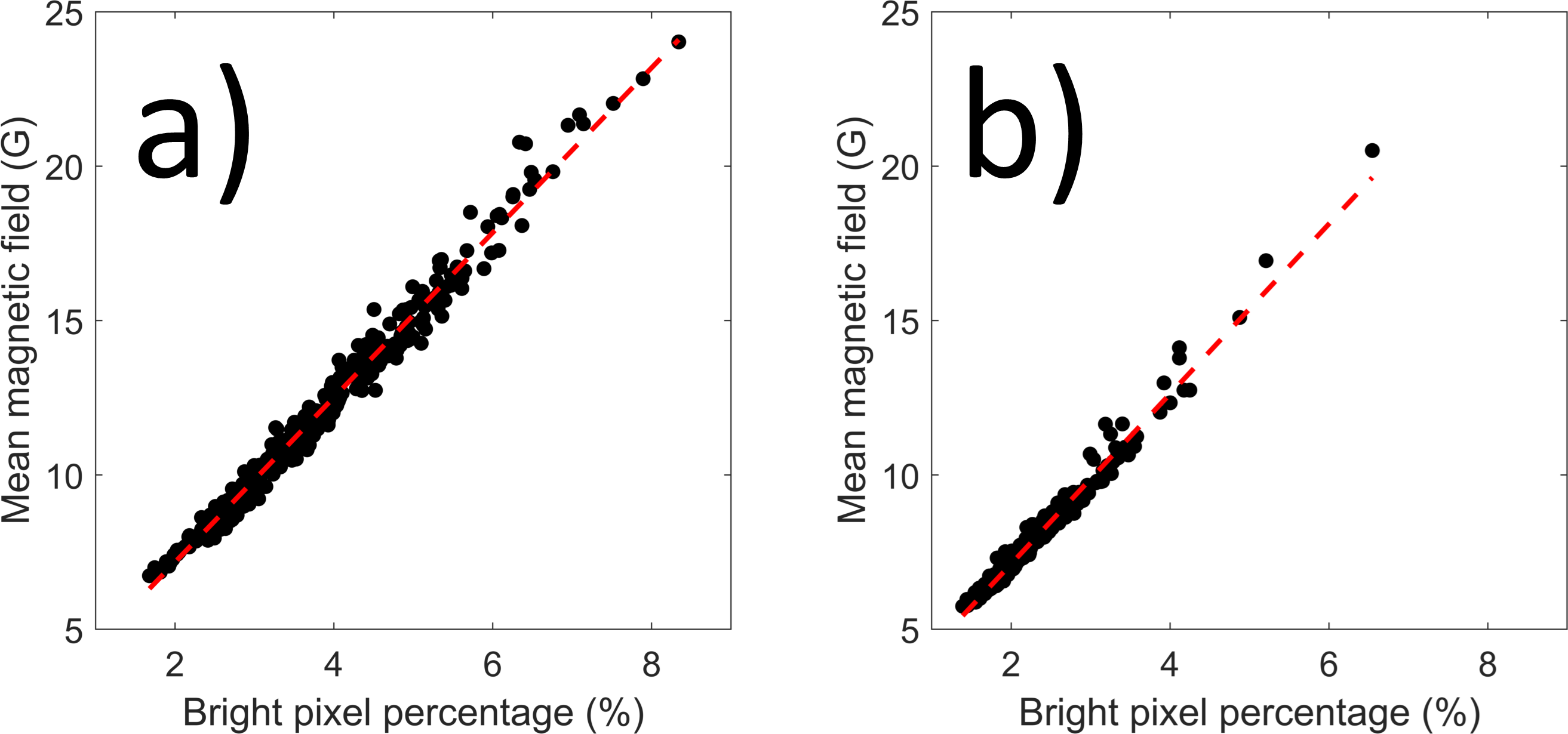}
   \caption{Mean magnetic field as a function of bright pixel percentage for the time a) before and b) after the  HMI measurement scheme change on 13 April 2016. Only days with a dark pixel percentage less than 0.001\% are shown. The red lines show the best least-squares regression lines. The best-fit parameters are shown in Table \ref{table:B_vs_AP}. \label{B_vs_BP_Fit}} %
    \end{figure*}

Figure \ref{BrightPixels}b shows the mean unsigned magnetic field as a function of the bright pixel percentage.
Similarly to in Fig. \ref{MeanAIAvsHMI}, there is some variability in the bright pixel percentage--mean unsigned magnetic field relation, which is closely related to the dark pixel percentage.
Days with the dark pixel percentage of less than 0.001\% form almost a straight line, representing a mean magnetic field baseline.
The higher the dark pixel percentage  for a fixed bright pixel percentage, the larger the magnetic intensities.
Days with a large magnetic field seem to be more typical when the dark pixel percentage increases.  
Figure \ref{B_vs_BP_Fit} shows the same relation as Fig. \ref{BrightPixels}b for those days when the dark pixel percentage is less than 0.001\%.
Panel a shows the relation for the time before the  HMI measurement scheme change and panel b for the time after the measurement scheme change. 
The related regression coefficients are shown in Table \ref{table:B_vs_AP}.
Again, the mean deviation is almost twice as small for the later period (MAD~=~0.186) than for the earlier period (MAD~=~0.334), indicating a more linear relation from the HMI scheme change.
The increase in bright pixel percentage of 1~ percentage point (pp) is associated with a roughly 2.7 G increase in the mean unsigned magnetic field. 
We note that that $R^2$ is slightly higher for the earlier period due to larger range of magnetic field values.

    \begin{table*}
\caption{Regression coefficients of the model $\langle B \rangle~=~\beta_{0} + \beta_{1}f_{BP}$.} 
\label{table:B_vs_AP} 
\centering 

\begin{tabular}{c|c c|c c} 
\hline\hline 
Time & $\beta_0$ & $\beta_1$ & $R^2$ & MAD \\ 
\hline 

Before  &   1.84$\pm$0.07 &   2.67$\pm$0.02     &   0.983   &   0.334 \\
After   &   1.62$\pm$0.06 &   2.75$\pm$0.02     &   0.979   &   0.186 \\

\hline
\end{tabular}
\end{table*}

\begin{figure*}[hbt!]
   \centering
  \includegraphics[width=\textwidth,height=\textheight,keepaspectratio]{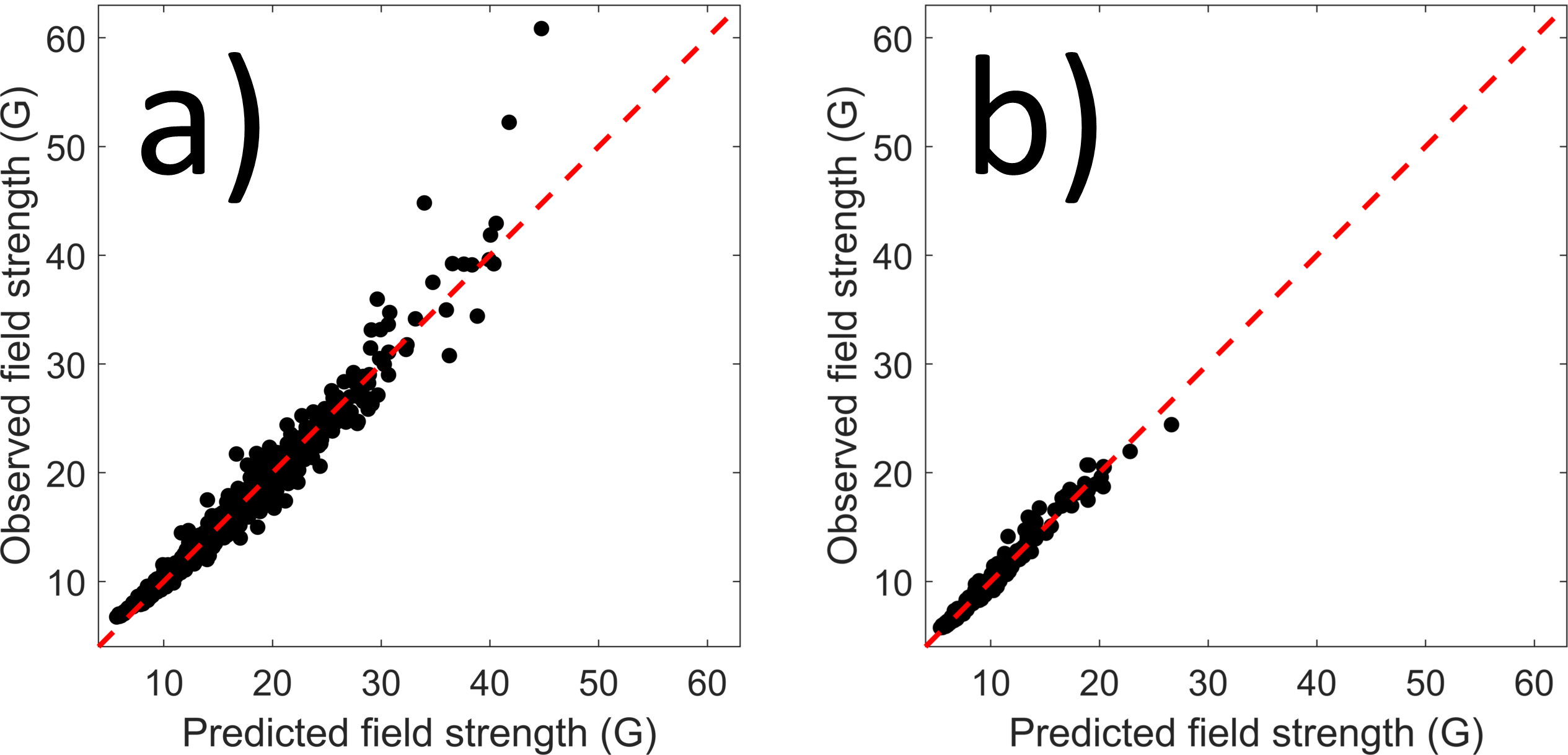}
   \caption{Scatter plot of observed mean unsigned magnetic field as a function mean unsigned magnetic field predicted from the regression model a) before and b) after the HMI measurement scheme change on 13 April 2016. \label{fig:regression}}%
\end{figure*}

The observation in Fig. \ref{BrightPixels}b that the mean unsigned magnetic field is dependent on both bright pixel percentage and dark pixel percentage motivated us to develop a multilinear regression model for the mean unsigned magnetic field
$$ 
\langle B \rangle~=~\beta_{0} + \beta_{1}f_{BP} + \beta_{2}f_{DP},
$$
where $f_{BP}$ is the bright pixel percentage and $f_{DP}$ is the dark pixel percentage.
We calculated this regression separately before and after the HMI measurement scheme change.
The regression results are shown in Fig. \ref{fig:regression} and Table \ref{table:regression}.
Figure \ref{fig:regression} shows the scatter plot of the observed mean magnetic field as a function of the mean magnetic field predicted from the regression model.
For this two-parameter regression model, 83.4\% of predictions are within 1~G of the observed value and 95.3\% within 2 G.
The standard deviation of residuals is 1.07 G during the earlier period and 0.37 G during the later.
The values of $\beta_1$, the coefficient of the bright pixel percentage, change only slightly between the two periods.
There is a greater change in $\beta_2$, representing the effect of the dark pixels.
In the earlier period, a 1 pp change in the dark pixel percentage is associated with a change of 42 G in the mean unsigned magnetic field.
In contrast, in the later period, a 1 pp change is associated with a change of 30 G.
The mean deviation of residuals is again much smaller in the later than the earlier period.

\begin{table*}[hbt!]
\caption{Regression coefficients of the model $\langle B \rangle~=~\beta_{0} + \beta_{1}f_{BP} + \beta_{2}f_{DP}$.} 
\label{table:regression} 
\centering 

\begin{tabular}{c|c c c|c c} 
\hline\hline 
Time & $\beta_0$ & $\beta_1$ & $\beta_2$ & $R^2$ & MAD \\ 
\hline 
Before  &   0.69$\pm$0.14       &   2.96$\pm$0.03       &   42.38$\pm$1.03   & 0.960   & 0.783   \\
After   &   1.32$\pm$0.09       &   2.92$\pm$0.04   &   30.21$\pm$0.69   & 0.976   & 0.329   \\
\hline
\end{tabular}
\end{table*}

\section{Bright and dark clusters}
\label{sec:clusters}
We used bright pixels (I $\geq$ 1.95, see Fig. \ref{fig:Histograms}) to construct bright clusters as 4-connected regions of bright pixels in every $0.4R_{\sun}$ solar disk image.
Similarly, dark clusters are defined as 4-connected regions of dark (I~<~0.5) pixels.
We searched for bright and dark clusters separately for every image, and did not track their temporal evolution from day to day.
We excluded all clusters smaller than ten pixels.
Above ten pixels the distribution of cluster sizes follows a power law, but below ten pixels it levels off to a normal distribution, thus representing different statistics.
A bright AIA 1600~\AA{} cluster is a radiative structure that includes different emissive structures from the granulation scale up to large-scale structures such as plages and sunspots.
In units of a millionth of the solar disk (MSD), which we use hereafter when reporting on cluster sizes, ten pixels equals about one MSD.
The final AIA 1600~\AA{} cluster dataset consists of 954\,028 bright clusters and 1328 dark clusters.
The size of bright clusters ranges from 1 MSD to 4137 MSD.
The median size of bright clusters is only 2.2 MSD (about 2080 km), so most of them have only a slightly larger size than granulation.
In comparison, the median size of dark clusters is 12.4 MSD (about 4900 km), while their size ranges from 1 MSD to 577 MSD.

\subsection{Bright clusters}
The bright cluster dataset includes  information on the size, the mean unsigned magnetic field, and the mean 1600~\AA{} contrast of each individual bright cluster observed within 0.4 $R_{\sun}$ between 1 March 2014 and 9 June 2017.

\subsubsection{Mean magnetic field versus size}
\begin{figure*}[ht]
 \centering
 \includegraphics[width=\textwidth,height=\textheight,keepaspectratio]{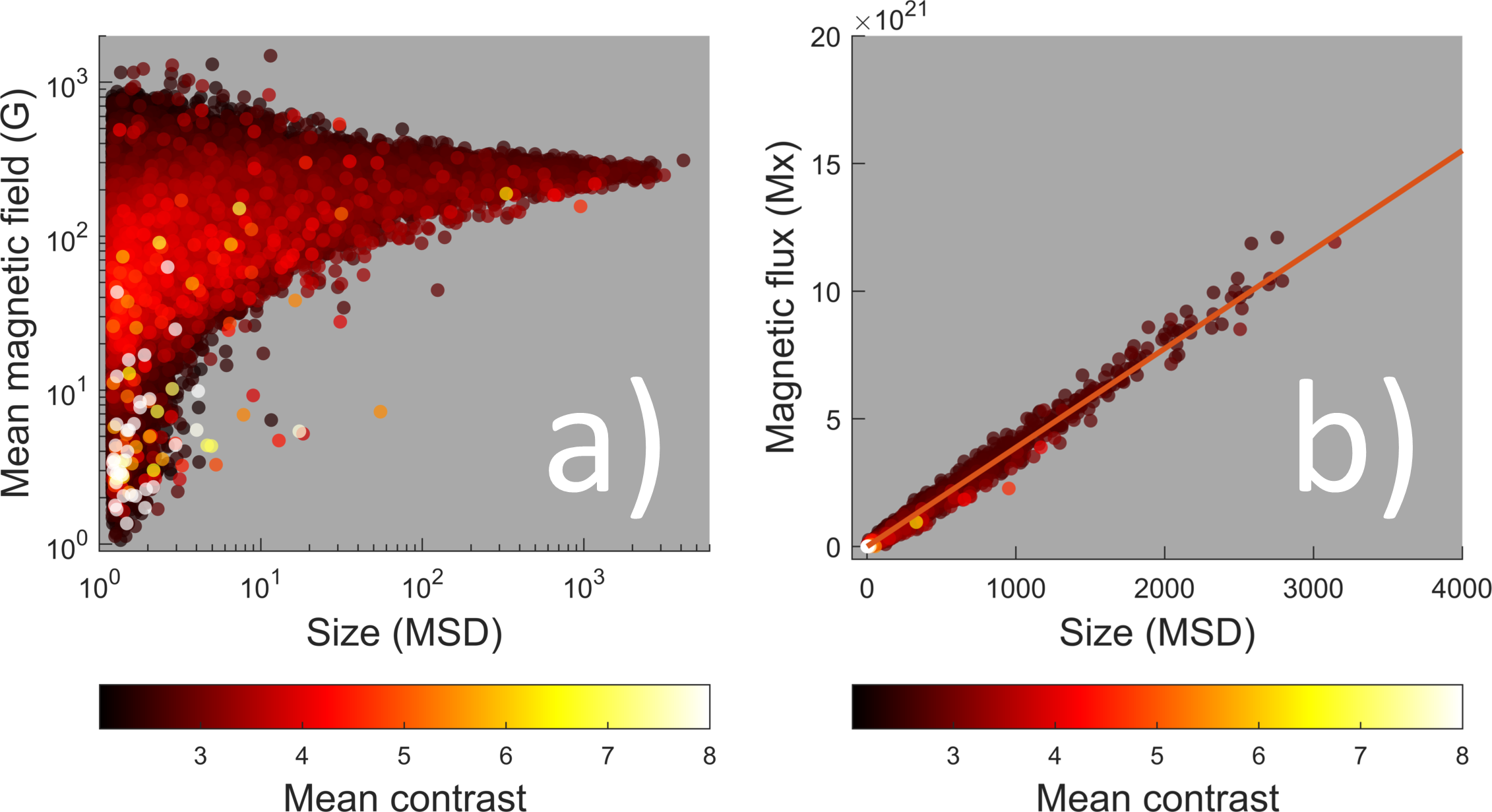}
 \caption{Scatter plots of a) the mean unsigned magnetic field of a bright cluster vs. its size and  b) the total unsigned flux of a bright cluster vs. its size. The color shows the mean 1600~\AA{} contrast of the bright cluster. \label{B_vs_S}}
\end{figure*}

Figure \ref{B_vs_S} shows the scatter plot between the mean unsigned magnetic field and the size of bright clusters.
The color of the dots denotes the mean 1600~\AA{} contrast in the corresponding bright cluster.
The color scale in Fig. \ref{B_vs_S} is set to saturate at I~=~8 because, even though the contrast of bright clusters ranges up to I~=~55, there are only 33 clusters whose contrast exceeds I~=~8.
Figure \ref{B_vs_S} shows that the smallest bright clusters of about 1 -- 2 MSD may have any value of mean unsigned magnetic field from practically 0~G to above 1000 G. 
When the size increases, the mean unsigned magnetic field starts to concentrate to a  narrower range, which is rather tightly bounded  from above and from below.
The average unsigned magnetic field of 4546 bright clusters with 100 MSD or more is 244 G with a standard deviation of only 33 G.

It should be noted that the bright clusters with the most intense mean magnetic field of B~>~500~G are fairly small, smaller than 10 MSD, typically only a few MSD.
These clusters are probably newly emerged magnetic flux tubes that cover only a small area, but have very strong magnetic field.
On the other hand, most of the brightest clusters seem to collect into a small region   only a few MSD in size   and with only a weak or moderate unsigned magnetic field of less than 100 G.
The median size of the brightest (I~>~5) clusters is only 1.5 MSD, and the median mean magnetic field B~=~5 G.

Figure \ref{B_vs_S}b shows the total (unsigned) magnetic flux within a bright cluster as a function of its size.
This relation is close to linear and the unsigned flux $\mathrm{\Phi}$ of a bright cluster can be expressed as
\begin{equation}
   \mathrm{\Phi}~=~(254.7*N_{\mathrm{pxl}} - 3386)\mathrm{G}\times{}\mathrm{A_{pxl}}, \\
\end{equation}
where $N_{\mathrm{pxl}}$ is the size of the cluster in pixels and $\mathrm{A_{pxl}}~=~435$ $\mathrm{km}^2$ is the area of a single pixel. 
The slope of the regression line is 254.7$\pm0.1$~G, which is an alternative way to determine the mean magnetic field of bright clusters.
This is how the mean magnetic field of plages was first calculated by \citet{Schrijver1987} and how it has been calculated more recently in an effort to reconstruct the past magnetic activity based on Ca II K plages (see, e.g., \citealt{Virtanen2019a}).

\subsubsection{Mean 1600~\AA{} contrast versus size}
\begin{figure*}[ht]
 \centering
 \includegraphics[width=\textwidth,height=\textheight,keepaspectratio]{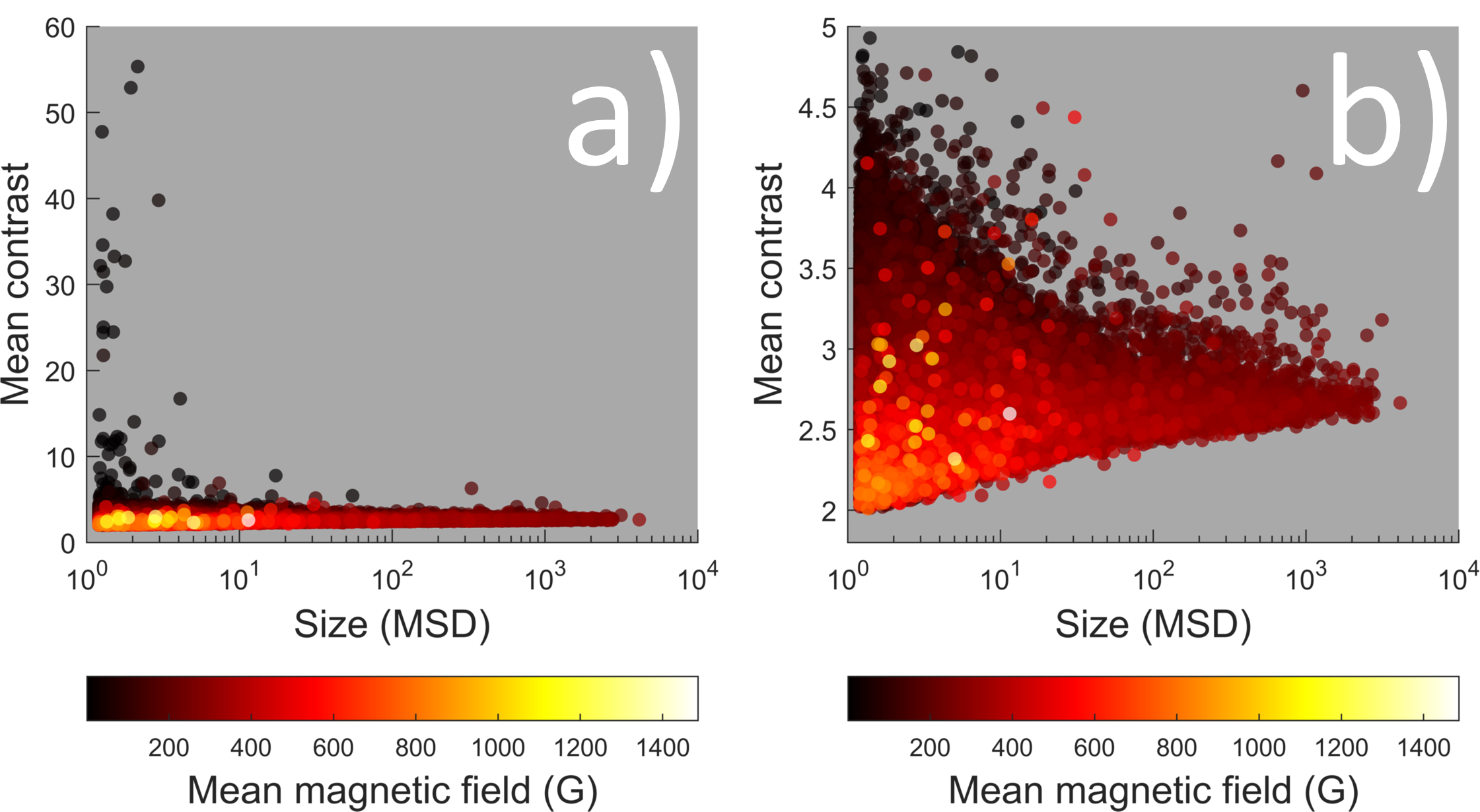}
 \caption{Scatter plots of the mean AIA 1600~\AA{} contrast of a bright cluster vs. its size. a) All clusters. b) Clusters with the mean contrast less than 5. The color shows the mean unsigned magnetic field of the bright cluster. \label{I_vs_S}}
\end{figure*}

Figure \ref{I_vs_S}a shows the scatter plot between the mean contrast and the size of bright clusters. 
The color of the dots now denotes the mean magnetic field of the corresponding bright cluster.
Some 20--30 small, very intense bright clusters with small mean magnetic field dominate the vertical range of Fig. \ref{I_vs_S}a.
These clusters correspond to the brightest data points in Fig. \ref{B_vs_S}a.
These brightest clusters are likely caused by energetic particles hitting the AIA 1600~\AA{} CCD detector.
The AIA 1600~\AA{} and 1700 \AA{} FUV channels have not been de-spiked, unlike the  extreme UV channels (for a detailed explanation, see the Guide to SDO Data Analysis\footnote{\url{https://www.lmsal.com/sdodocs/doc/dcur/SDOD0060.zip/zip/entry/}}).

Figure \ref{I_vs_S}b shows the same plot as Fig. \ref{I_vs_S}a, but with a more limited y-axis range that better shows the bulk of (true) bright clusters.
For all cluster sizes, the mean contrast range is tightly bound from below, and the lower limit increases with size fairly systematically.
The upper mean contrast limit is more scattered but decreases with size on average.
The average mean contrast of the 4546 bright clusters with 100 MSD or more is 2.67, with a standard deviation of only 0.13.
Bright clusters with the strongest magnetic field lie in the lower left corner of this scatter plot.
They are rather small (<~10 MSD) and have at most moderate contrast (I~<~3).
Figure \ref{I_vs_S}b resembles Fig. \ref{B_vs_S}a in the sense that the mean contrast converges to an increasingly narrow range when the size increases,  as the mean unsigned magnetic field does in Fig. \ref{B_vs_S}a.

\begin{figure*}[ht]
 \centering
 \includegraphics[width=\textwidth,height=\textheight,keepaspectratio]{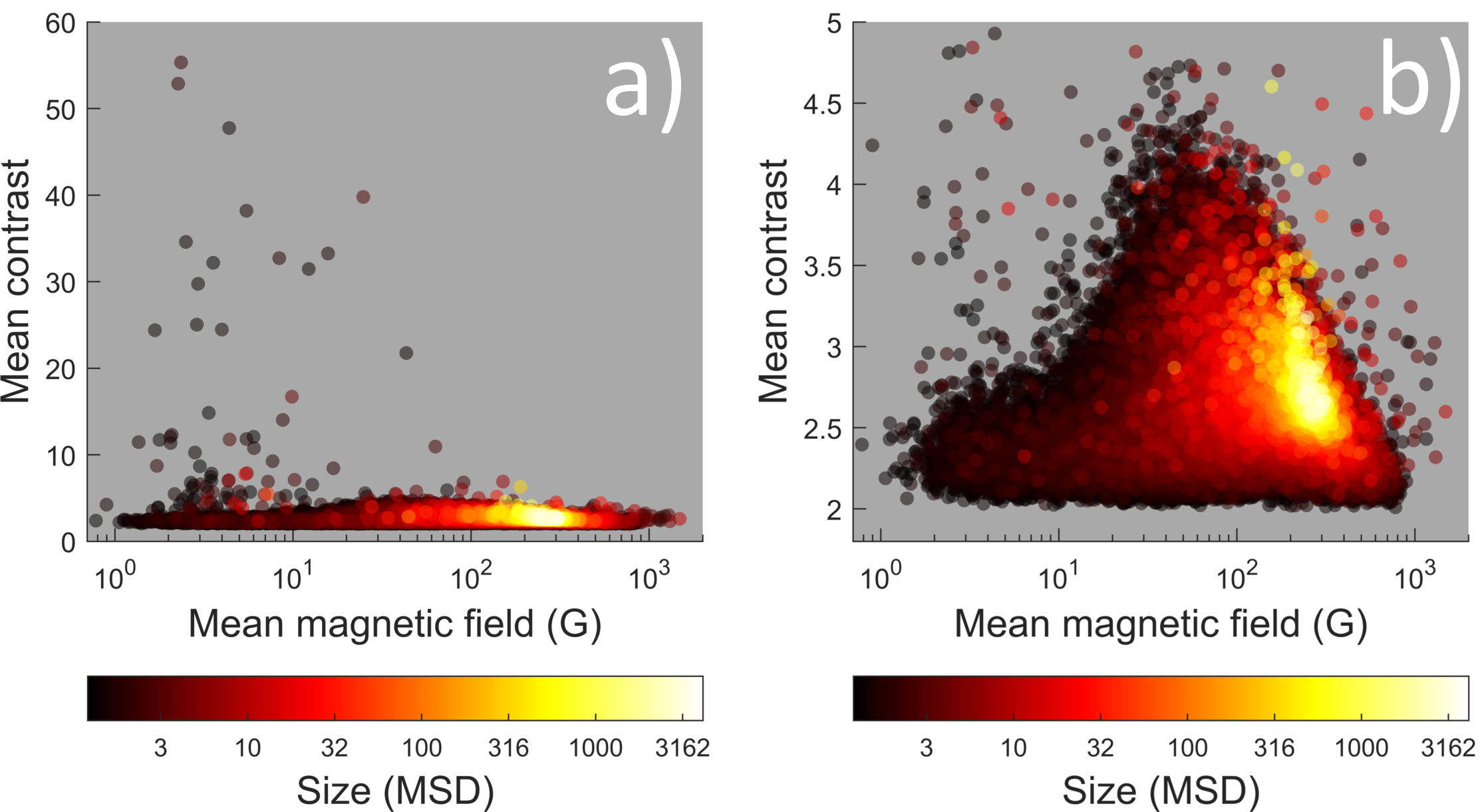}
 \caption{Scatter plot of the mean AIA 1600~\AA{} contrast of a bright cluster vs. its mean unsigned magnetic field. a) All clusters. b) Clusters with the mean contrast less than 5. The color shows the cluster size.  \label{I_vs_B}}
\end{figure*}

\subsubsection{Mean 1600~\AA{} contrast versus mean magnetic field}
Figure \ref{I_vs_B}a shows the scatter plot between the mean contrast and the mean unsigned magnetic field of bright clusters. 
The color of the dots indicates now the size of the bright cluster.
Again, the brightest clusters due to the likely false signals dominate the vertical range of Fig. \ref{I_vs_B}a as they did in Fig. \ref{I_vs_S}a.
Figure \ref{I_vs_B}b shows the same data as Fig. \ref{I_vs_B}a, but with a limited y-axis range.
The scatter plot of Fig. \ref{I_vs_B}b for the bulk of bright clusters almost resembles  a triangle.
For a given mean magnetic field, the mean AIA 1600~\AA{} contrast of a bright cluster can have a range of values, whose lower bound is almost constant and whose upper bound depends on the mean magnetic field.
Until about B~=~50~G, this upper bound roughly increases with the magnetic field, and above it this limit decreases.
The largest bright clusters lie on the right-hand side of the triangle, not at the largest mean magnetic field values, but with the mean unsigned magnetic field of around 100 -- 300~G, in agreement with Fig. \ref{B_vs_S}a.

\subsection{Dark clusters}
\begin{figure*}[ht]
 \centering
 \includegraphics[width=\textwidth,height=\textheight,keepaspectratio]{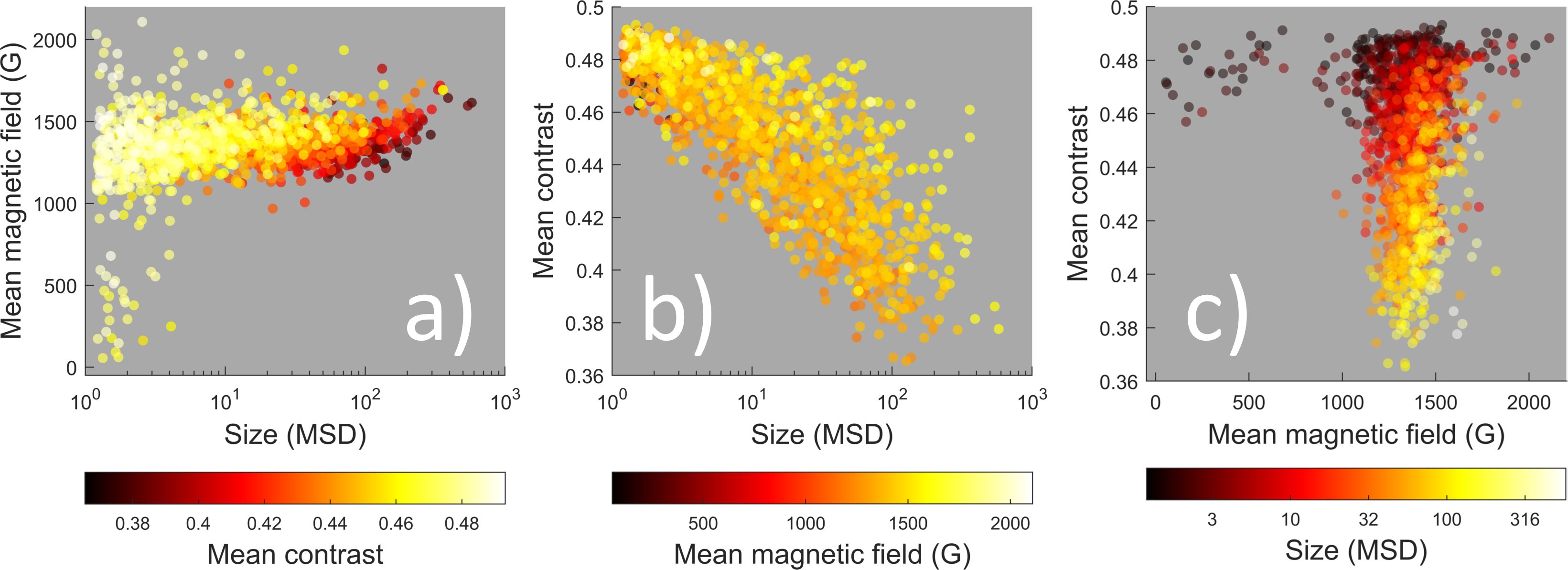}
 \caption{Dark cluster relations. a) Mean unsigned magnetic field vs. size. The color shows the mean 1600~\AA{} contrast of the cluster.
 b) Mean AIA 1600~\AA{} contrast vs. size. The color shows the mean unsigned magnetic field of the cluster.
 c) Mean AIA 1600~\AA{} contrast vs. mean unsigned magnetic field. The color shows the cluster size.
 \label{fig:DarkClusters}}
\end{figure*}

The dark cluster dataset, similarly to the bright cluster dataset, includes the information on the size, the mean magnetic field, and the mean 1600~\AA{} contrast of each dark cluster observed within 0.4$R_{\sun}$ between 1 March 2014 and 9 June 2017.
Figure \ref{fig:DarkClusters} shows the same relations between the properties of dark clusters as those presented for bright clusters in Figs. \ref{B_vs_S}a, \ref{I_vs_S}a, and \ref{I_vs_B}a.
Figure \ref{fig:DarkClusters}a shows the scatter plot between the mean unsigned magnetic field and the size of dark clusters.
Most field values are independent of cluster size and have field values between 1.0--1.8~kG.
They are approximately normally distributed around the mean of $\mu$~=~1371 G with a standard deviation $\sigma$~=~126~G.
The mean contrast of the dark cluster decreases with the cluster size, as seen in Fig. \ref{fig:DarkClusters}b.
This inverse relation is also visible in Fig. \ref{fig:DarkClusters}a, where the largest dark clusters are colored darker than the smaller clusters.
However, the mean contrast does not depend on the mean magnetic field, as seen in Fig. \ref{fig:DarkClusters}c.
This is also visible in Fig. \ref{fig:DarkClusters}b, where no difference in color is found along the scatter plot.

\section{Relevance for Ca II K-based reconstructions}
\label{sec:cak}
In this paper we study the relation between AIA 1600~\AA{} contrast and HMI magnetic fields.
However, these results are also useful for magnetic field reconstructions based on other wavelengths, for example  those obtained from Ca II K spectroheliograms.
Earlier, \citet{Rutten1999} and \citet{Loukitcheva2009}  studied the relation between Ca II K and TRACE 1600~\AA{} emissions whose spectral content is close to that of AIA 1600~\AA{}. 
\citet{Rutten1999} showed that the bright features observed by TRACE 1600~\AA{} correspond closely to features seen in Ca II K observations taken at the Swedish Vacuum Solar Telescope on La Palma, while \citet{Loukitcheva2009} showed a linear relation between TRACE 1600~\AA{} and Ca II K brightness as measured by the  Big Bear Solar Observatory.
More recently, \citet{Bose2018} compared the total areas of bright features in co-temporal AIA 1600~\AA{} and Ca II K images taken by the Chromosperic Telescope \citep[ChroTel;][]{Bethge2011} at the Observatorio del Teide in Izaña on Tenerife.
They found that the total areas matched well, the difference being less than 1.5 \%.

We studied the relation between AIA 1600~\AA{} and Ca II K emissions, in 115 co-temporal AIA 1600~\AA{} and Ca II K observations made by ChroTel.
We calibrated the ChroTel images with the same method that we used for the  AIA data.
Since ChroTel observations have a slightly lower resolution of 1.029 arcsec/pxl, we downgraded the  AIA images to the same resolution.
We also applied a Gaussian filter to the AIA images with different window sizes in order to mimic the effect of the atmosphere to ChroTel observations.

\begin{figure*}[ht]
 \centering
 \includegraphics[width=\textwidth,height=\textheight,keepaspectratio]{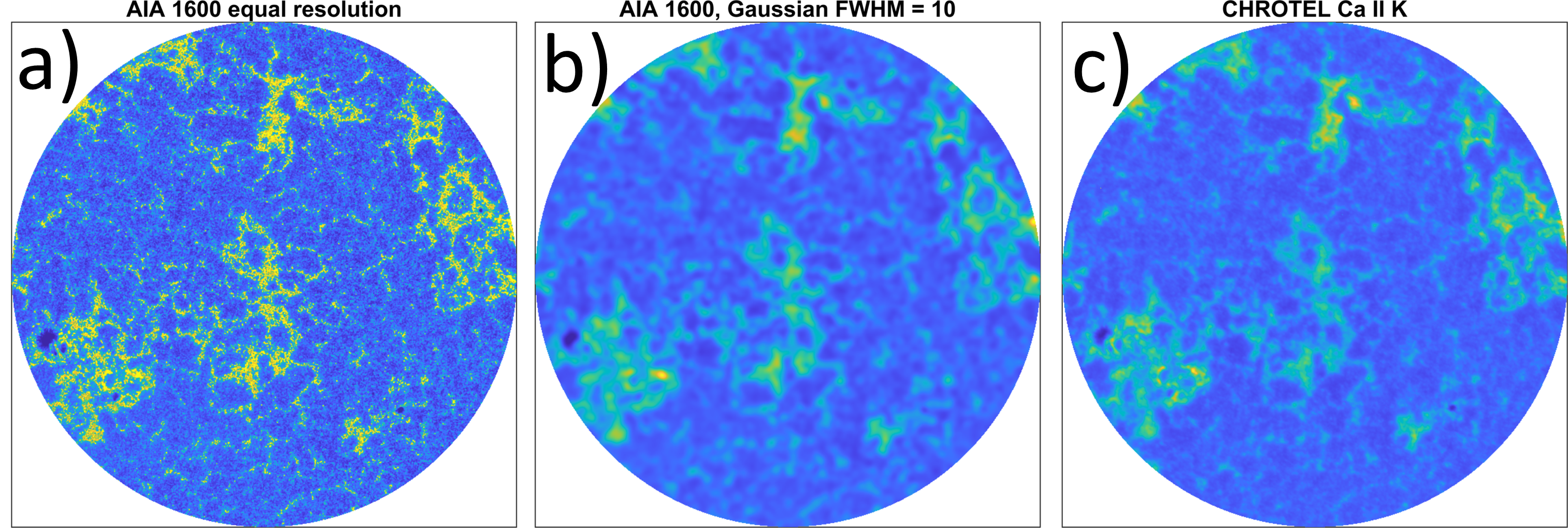}
 \caption{Co-temporal AIA 1600~\AA{} and ChroTel Ca II K images of the $0.4R_{\sun}$ disk taken on 1 April 2014 at 10:00 UTC. a) AIA 1600~\AA{}; b) blurred AIA 1600~\AA{} (FWHM~=~10 pixels); c) ChroTel Ca II K.
 \label{fig:AIACakPics}}
\end{figure*}

\begin{figure*}[ht]
 \centering
 \includegraphics[width=\textwidth,height=\textheight,keepaspectratio]{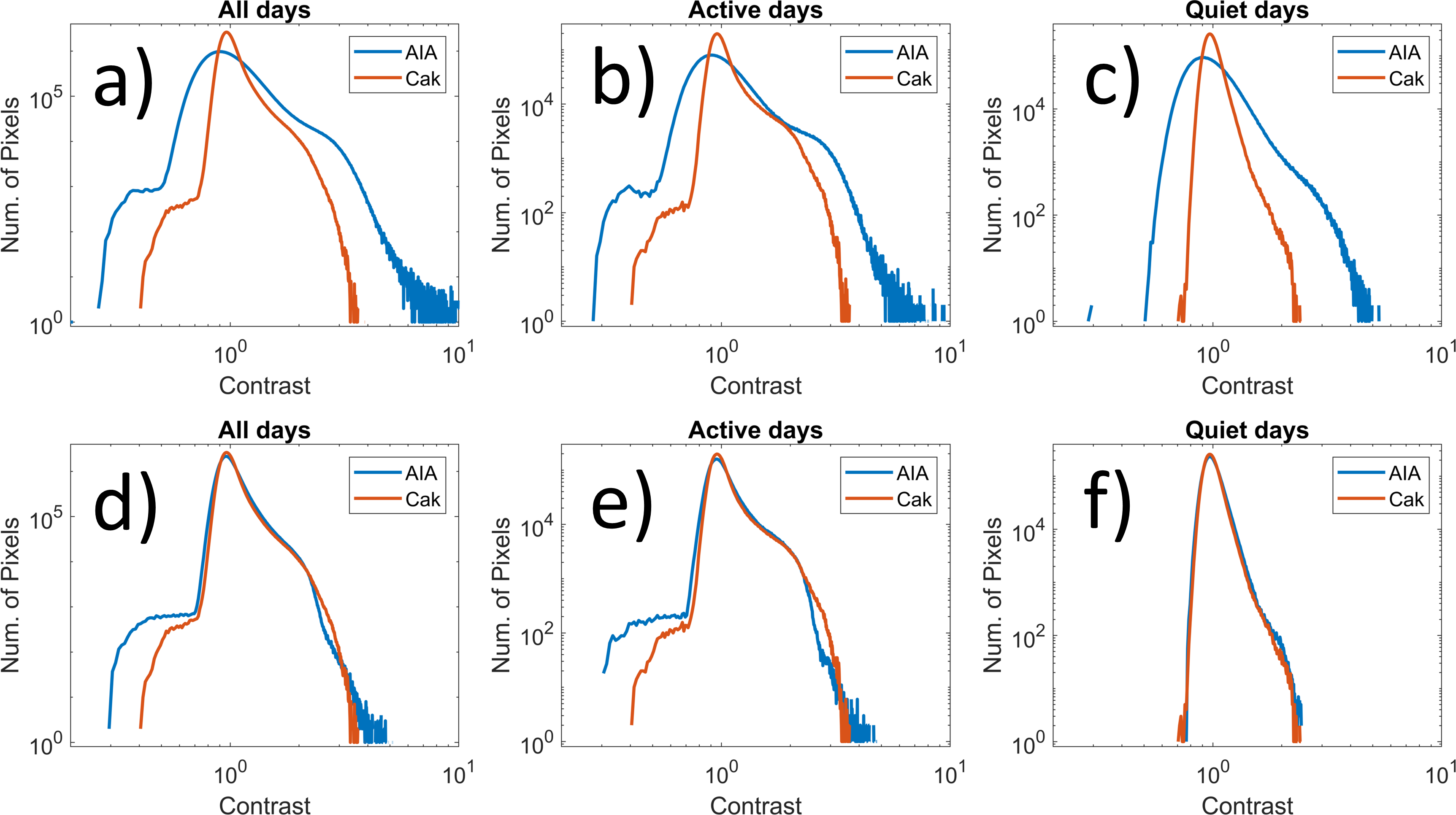}
 \caption{Pixel contrast distributions of AIA 1600~\AA{} (blue) and ChroTel Ca II K (orange). Top row: Unblurred AIA 1600~\AA{}. Bottom Row: Blurred AIA 1600~\AA{} (FWHM~=~10 pixels). ChroTel Ca II K distributions are the same for both rows. Left column: All 115 days. Middle column: The 10 most active days. Right column: The 10   quietest days.
 \label{fig:AIACakDists}}
\end{figure*}

Figure \ref{fig:AIACakPics} shows the co-temporal  AIA 1600~\AA{} image, the  blurred AIA 1600~\AA{} image,  and the  ChroTel Ca II K image of $0.4R_{\sun}$ disk taken on 1 April 2014 at 10:00 UTC.
Figure \ref{fig:AIACakDists} shows the pixel contrast distributions of simultaneous AIA-ChroTel observations for all 115 days (panels a and d), for the 10 most active days (panels b and e), and for the 10  quietest days (panels c and f).
The upper row shows the distributions for the unblurred AIA images and the lower row for the blurred AIA images.
Simultaneous Ca II K distributions are the same for both rows.
Although the AIA 1600~\AA{} image in Fig. \ref{fig:AIACakPics}a has nominally the same resolution as the ChroTel image in Fig. \ref{fig:AIACakPics}c, the AIA image displays much more detail than ChroTel, mainly due to the scattering and absorption of light in the Earth's atmosphere, which effectively blurs the ChroTel image.
From the contrast distributions in the upper row of Fig. \ref{fig:AIACakDists}, we see that the dynamic range of ChroTel is much narrower than that of AIA.
Thus, AIA is much more sensitive, especially at the low end of the contrast spectrum.

Figure \ref{fig:AIACakPics}b depicts the same image as Fig. \ref{fig:AIACakPics}a, but blurred with a Gaussian window whose full-width at half maximum (FWHM) is 10 pixels.
The pixel contrast distributions of similarly blurred AIA 1600~\AA{} images for all 115 days (panel d), for the 10 most active days  (panel e), and for the  10  quietest days (panel f) are shown in the lower row of Fig. \ref{fig:AIACakDists} together with ChroTel Ca II K pixel contrast distributions.
When using the 10-pixel Gaussian, the blurred AIA 1600~\AA{} image (in Fig. \ref{fig:AIACakPics}b) and the ChroTel Ca II K image (in Fig. \ref{fig:AIACakPics}c), as well as the corresponding pixel distributions in Fig. \ref{fig:AIACakDists}, resemble each other very closely.
The quiet-day distributions almost overlap, but the active-day distributions still show the slightly weaker sensitivity and greater saturation of ChroTel.
Considering that AIA and ChroTel are quite different instruments and that simple blurring by Gaussian filtering brings contrast distributions so close to each other, this gives strong evidence for a close connection between AIA 1600~\AA{} contrast and Ca II K emission.

\section{Discussion}
\label{sec:disc}
In this paper we   introduced the concepts of bright and dark pixels and bright and dark clusters using the AIA 1600~\AA{} contrast, and   studied the relation between average contrast (I) and unsigned magnetic flux (B) and size of clusters.
Bright (dark, resp.) clusters are 4-connected regions of bright (dark) pixels. 
Bright and dark pixels are associated with magnetic populations of moderate and strong field pixels, respectively, obtained from HMI magnetograms. 
Dark pixels were defined as   pixels whose contrast is below I~=~0.5 and bright pixels as those whose contrast is greater than I~=~1.95.
The bright pixel threshold corresponds to the contrast level that maximizes the average size of bright clusters.
The dark pixel upper limit corresponds to the contrast level, below which there are practically no pixels on quiet days.
We showed that the average size of magnetic clusters displays a bimodal distribution as a function of threshold.
The peaks of this distribution correspond to the moderate field (B~=~55 G) and strong field (B~=~1365~G) magnetic threshold values, which maximize the average size of these two populations of magnetic clusters and produce in this sense the most coherent clusters within these two populations.
The value we found for the moderate field is in agreement with that of  \citet{Schrijver1987}, who found that B~=~50~G produced the most coherent active regions, which he called magnetic plages.
Our AIA 1600~\AA{} contrast thresholds can be considered optimal since they correspond to these magnetic thresholds (moderate field $\Leftrightarrow{}$ bright pixels, strong field $\Leftrightarrow{}$ dark pixels), and thus have a clear physical meaning.

\subsection{Disk averages}
The disk-averaged (using $0.4 R_\sun{}$ disk) AIA 1600~\AA{} contrast increases roughly linearly with the corresponding mean unsigned magnetic field, but there is wide variability that increases with increasing magnetic field (see Fig. \ref{MeanAIAvsHMI}).
This variability depicts a saturation pattern that is clearly organized by the percentage of dark pixels present on the solar surface.
A fairly small percentage of dark pixels of less than 0.1\% is sufficient to deviate the respective day from the linear relation.
Figure \ref{MeanAIAvsHMIFit} shows that the relation between the mean contrast and mean unsigned magnetic field becomes almost linear when the dark pixel percentage is less than 0.001\%.

Saturation of chromospheric emission with increasing magnetic field has been noted earlier in both disk-averaged and pixel-by-pixel relations.
\citet{Schrijver_1989} suggested that the saturation can be explained in terms of flux tube geometry, assuming that the radiative losses of flux tubes are proportional to their area at the emission height.
Since expanding nearby flux tubes meet at some altitude, the sum of their projected areas above this merging height is less than that of individual flux tubes at the same altitude.
If the merging of flux tubes happens below the emission height, their emission is less than that of individual flux tubes, which causes the saturation.
The qualitative aspects of this picture were confirmed by the two-dimensional model of \citet{Solanki1991}.

We found that the percentage of bright pixels present on the solar disk almost entirely explains the variability of the mean contrast ($R^2$~=~0.998, see Fig. \ref{BrightPixels}a).
There is a robust linear relationship between these two parameters that  is not affected by the dark pixel percentage.
Thus, it also holds on those days when the mean contrast is saturated due to a large unsigned mean magnetic field (see Fig. \ref{MeanAIAvsHMI}).
Since this relation between mean contrast and bright pixel percentage also holds on days when the dark pixel percentage is large, the excess of dark pixels (e.g., sunspots) decreases the mean contrast mainly by reducing the number of the bright pixels, not by the low contrast of dark pixels.

A situation analogous to Fig. \ref{MeanAIAvsHMI} is found in Fig. \ref{BrightPixels}b, which depicts the relation between the mean unsigned magnetic field and bright pixel percentage.
While most points are quite closely linearly related, there are points above the average regression (with larger magnetic field), which have an increasingly large dark pixel percentage.
We note that  the points with large percentage of bright pixels  also have a large percentage of dark pixels.
This is natural since sunspots occur together with other active regions.
Figure \ref{B_vs_BP_Fit} shows that when the dark pixel percentage is small, this nonlinearity disappears, and the bright pixel percentage has a strictly linear relation with the mean unsigned magnetic field (see also Table \ref{table:B_vs_AP}).
When the dark pixel percentage is larger, this relation gives the minimum of the mean unsigned magnetic field for a fixed bright pixel percentage.
When we use the dark pixel percentage with the bright pixel percentage in a multilinear regression, we obtain a better estimate for the mean unsigned magnetic field over the whole range of values (see Fig. \ref{fig:regression} and Table \ref{table:regression}).
We find that 95\% of the values predicted by this model lie within $\pm$ 2~G of the actual values.
This result allows us to express the magnitude of the global mean magnetic field in terms of the percentages of bright and dark pixels, which could be useful when constructing the past solar magnetic activity.
Since the  1600~\AA\ contrast has a close spatial correspondence \citep[][see also Fig. \ref{fig:AIACakDists}]{Rutten1999,Bose2018} and is linearly related to Ca II K emission \citep{Loukitcheva2009}, it is reasonable to expect that the bright and dark pixel percentages are related to fractional areas of Ca II K plages and sunspots;  there are datasets for both that cover   more than a century.
However, a more careful analysis between the  AIA 1600~\AA\ and Ca II K data is needed and is the subject of future work.

The analysis of disk-averaged data shows a slight change in the relation between contrast and unsigned magnetic field after HMI changed the scheme for collecting vector magnetic field measurements.
This change decreased the noise level in magnetic field measurements, which is important mainly for weak fields.
Stronger fields are relatively less affected by noise or its reduction (see Fig. \ref{fig:HMINoise}).
We note that none of our main conclusions are affected by this change, although some regression parameters are slightly different before and after the HMI measurement scheme change.

\subsection{Bright clusters}
We found that the bright clusters have the mean unsigned magnetic field B~=~254.7 $\pm$ 0.1~G (see Fig. \ref{B_vs_S}b).
As the size of bright clusters grows, the mean unsigned magnetic field converges towards this constant value (Fig. \ref{B_vs_S}a).
\citet{Schrijver1987}, who used observations of Kitt Peak 40-channel magnetograph, found that the magnetic field of a plage is nearly constant and about $100 \pm 20$~G.
Later, using a larger sample of active regions, \citet{Schrijver1994} found a somewhat higher flux density of about B~=~135--150~G for plages.
Similar average magnetic fields for plages have recently been used to reconstruct the past magnetic activity of the Sun.
For example, \citet{Virtanen2019a}, who used synoptic maps constructed from observations of the  Kitt Peak 512-channel magnetograph, found the mean magnetic field of plages to be about 111 G.

The observation by \citet{Schrijver1987} that plages are radiative structures with a constant magnetic flux density suggests that the large AIA 1600~\AA{} bright clusters could be plages, or at least, be related to plages.
The fact that we find the mean magnetic field of bright clusters to be roughly 100~G larger than found earlier for plages is most likely due to the higher spatial resolution of the HMI instrument.
In lower-resolution observations the measured magnetic field is decreased both due to the mixing of fields of opposite polarities and due to mixing of weak and strong fields of the same polarity.
The former effect decreases the measured flux especially in the quiet Sun, but the latter effect is more prominent in the active regions since they are largely unipolar \citep{Krivova2004}.
If we compare the bright and magnetic clusters in our data (see Fig. \ref{fig:masks}) with plages, as determined by \citet{Chatzistergos2020a}, among others,  it is quite clear that our clusters are less uniform (less-connected, more fractal-like) than plages of  lower-resolution observations.
The difference between the high-resolution AIA 1600~\AA{} data and the low-resolution Ca II K data is also evident if we compare the  bright pixel lower threshold $I_{BP}~=~1.95$ to contrast levels in previous Ca II K studies.
It is well above the maximum contrast (about 1.5) of Mount Wilson Ca II K synoptic data \citep{Bertello2010,Pevtsov.etal2016,Virtanen2019a} and close to the maximum contrast of Precision Solar Photometric Telescope (Rome/PSPT) full-disk Ca II K data \citep{Chatzistergos2019b}.
Only a few studies explicitly mention the thresholds  used to distinguish different features, typically network and plages, from their data.
These thresholds vary from 1.1 to 1.35 \citep{Priyal2014,Priyal2019,Chatzistergos2019b,Singh2021}.
The higher resolution of AIA and HMI data allows a more accurate separation of bright and moderate to strong field pixels from the quiet Sun, which also leads to a larger mean magnetic field strength for large bright clusters and plages.
The question about the absolute level of the measured magnetic field has been studied by
\citet{Riley2014} and \citet{VirtanenIlpo2017}, among others,  who compared the magnetic field measurements between several solar observatories and showed large differences in magnetic field intensities between different datasets.

The constancy of plage mean magnetic field is an important fact, which allows a simple relation between plage area and its total magnetic flux.
Such a relation can be used in the reconstruction of magnetic fields, for example  using  historical Ca K line observations.
Since there is a good agreement between the 1600~\AA{} and Ca II K emissions, the constant mean magnetic field of bright clusters could be used to estimate the total magnetic field from the observed Ca II K plages \citep{Pevtsov.etal2016}.
This is important, for example, when predicting the early polar fields using the surface flux transport model and active regions which have a central role in the evolution of large-scale magnetic fields \citep{Virtanen2019a}.

We also found that the mean 1600~\AA{} contrast of the largest bright clusters (plages) is almost independent of cluster size (see Fig. \ref{I_vs_S}).
The average mean contrast of bright clusters larger than 100 MSD is I~=~ 2.67 $\pm$ 0.13.
This is in qualitative agreement with the results of  \citet{Schrijver1985}, who found that the plage mean contrast is virtually independent of plage size.

We found that the mean contrast of bright clusters as a function of the mean unsigned magnetic field has an upper bound that  depends on magnetic field, and a lower bound that is almost constant (see Fig. \ref{I_vs_B}).
While for weaker magnetic fields, the upper bound increases with increasing field, above approximately 50~G it starts to decrease.
The lower bound is set by the condition defining the bright pixels forming the bright cluster, but the upper bound has an interesting physical cause. It can be understood in terms of magnetic field energy input or production, for example  via reconnection within the bright cluster.
The contrast of a bright cluster would then indicate the amount of reconnection taking place within the cluster.
The larger the reconnecting flux inside the bright cluster, the more magnetic energy is released and transferred to thermal and bulk flow energy of chromospheric plasma, and the more background FUV radiation is produced, increasing the FUV contrast of the cluster.
The mean magnetic field of a bright cluster is a measure of how tightly packed the magnetic flux tubes within the cluster are and what is the fraction of the oppositely signed flux tubes.
Thus, the stronger the mean magnetic field of the cluster, the higher the probability of reconnection of oppositely signed flux tubes and the larger the FUV emission is.
So, as the mean magnetic field of a bright cluster grows, the reconnection rate first increases as the flux tubes are brought closer together, which is seen as the increasing upper bound of the bulk of bright clusters in Fig. \ref{I_vs_B}.

It should be noted that by far, the largest FUV contrasts are found for bright clusters with a rather weak magnetic field, mostly below 20~G (see Fig. \ref{I_vs_B}). 
This suggests that there are roughly an equal number of  flux tubes of either polarity in these clusters.
Moreover, as Fig. \ref{I_vs_B} also shows, the size of these clusters is very small, indicating that the oppositely signed flux tubes are densely packed in these clusters.
These clusters are probably related to polarity inversions, which have been connected with abnormally high Ca II K emission rates by \citet{Schrijver_1989} and \citet{HarveyWhite1999}.

The development of magnetic flux accumulation culminates as field strength approaches 50~G, beyond which the contrast upper bound turns to decrease with increasing magnetic field.
The mean magnetic field increases in these clusters that are dominated by flux tubes of the same polarity, where the signed mean of the magnetic field increases, while the probability of reconnection slowly decreases.
We also note   that the largest sizes of bright clusters are found in the middle of this range of decreasing contrast upper bound (see Fig. \ref{I_vs_B}b).
As also shown in Fig. \ref{B_vs_S}, all the largest bright clusters have roughly the same unsigned mean magnetic field of 254.7 G.
They also have roughly the same mean contrast of about 2.67.
This all suggests that the largest clusters have roughly the same density of magnetic flux tubes, and that the fraction of oppositely signed flux tubes is rather limited and roughly constant for all bright clusters.
These results give interesting new information about the main phase of plage evolution.

\subsection{Dark clusters}
The mean unsigned magnetic field of a dark cluster does not depend on its size (see Fig. \ref{fig:DarkClusters}a).
The mean magnetic field of dark clusters is approximately normally distributed around the mean $\mu$~=~1371 G with standard deviation $\sigma$~=~126 G.
These values match well with the known properties of sunspots whose typical magnetic field strength is 1000--1500~G \citep{Solanki2003}. 
While the mean unsigned magnetic field of dark clusters is roughly constant, their mean contrast is inversely proportional to the logarithm of their size (see Fig. \ref{fig:DarkClusters}b).
Although this inverse relation is quite systematic and statistically significant, the range of contrast variation is quite small.
Within 1.5 orders of magnitude of sizes from 3 to 100 MSD, the mean contrast decreases by only  20 \% from 0.48 to 0.40.
A similar relation was found by \citet{Mathew2007}, who showed that relation between the umbral contrast and umbral radius could be described by an inverse power law.
We also note that  the size of the smallest dark clusters (about 1550 km) is comparable to the size of pores (2000 -- 4000 km) and the umbra of the small sunspots \citep{Bogdan1988,Rucklidge1995}.

We also found that, as an interesting contrast to bright clusters, the mean contrast of dark clusters does not show any dependence on the mean magnetic field which is roughly constant (see Fig. \ref{fig:DarkClusters}c).
The variability in the contrast of the dark clusters comes only from their size, with smaller dark clusters being slightly brighter (see Fig. \ref{fig:DarkClusters}b).

\section{Conclusions}
\label{sec:conc}
In this paper we  studied the relation between the  SDO/AIA 1600~\AA{} emission contrast and unsigned photospheric magnetic fields from SDO/HMI, mainly within $0.4 R_{\sun}$-radius disk around the disk center.
We investigated this relation for bright and dark pixels and bright and dark clusters defined here in an objective and robust way. 
Our analysis indicates that the AIA 1600~\AA{} images are not consistently flat-fielded, and that the existing flat-fielding does not fully remove the long-term trend effect of detector degradation.
While studies using only a few 1600~\AA\ images might not be significantly affected by these calibration issues, the use of 1600~\AA\ brightness for long-term studies requires careful calibration and the removal of degradation effects.

We developed a robust and objective method to define two contrast thresholds that were used to define the bright (I $\geq$ 1.95) and dark (I~<~0.5) pixels from the bulk of more moderately bright pixels.
These contrast thresholds also correspond to magnetic field thresholds that produce maxima for the size distributions of the populations of moderate (55 G~<~B~<~475 G) and strong (B~>~1365~G) magnetic clusters obtained from HMI magnetograms.
Using these magnetic field thresholds the distribution of magnetic clusters is most coherent.
The bright clusters correspond to moderate field clusters and represent plages as well as the enhanced network (see Fig. 9).
The dark clusters correspond to strong field clusters and represent pores and sunspot umbrae.

We showed that the variability of AIA 1600~\AA\ contrast can be almost entirely explained by the variability of the number of bright pixels present on the solar surface.
This implies that the variation in the AIA 1600~\AA{} contrast over the solar cycle closely follows and can be represented by the change in   percentage of the solar disk occupied by bright pixels (typically forming plages and network), not by the change in their mean contrast.
While we cannot rule out that such changes in contrast may still occur on a smaller spatial scale, the observations at 0.6 arcsecond per pixel resolution do not show such changes.
We also showed that the bright pixel percentage serves as a good proxy of the disk mean unsigned magnetic field when the dark pixel percentage is close to zero.
When both bright and dark pixels are taken into account, we can fairly accurately estimate the mean unsigned magnetic field using a multilinear regression model with bright pixel percentage and dark pixel percentage as regression parameters.
We found that for 95\% of the daily images the regression model predicted the field strength within $\pm$ 2~G of the actual value.
This supports earlier results that historical datasets of sunspots and Ca II K plages can be used to produce a reliable proxy for the global solar unsigned magnetic field over more than one century.
Although we showed that blurred AIA 1600~\AA{} images correspond to ground-based Ca II K images very well, future work is needed to show in detail that this paper's results also apply to Ca II K data.

We found that the bright clusters have nearly a constant magnetic field of 254.7$\pm0.1$~G.
The large bright clusters correspond to chromospheric plages whose magnetic field is known to be constant.
However, the value we find is approximately 100~G larger than the field strength found in earlier studies and is used in reconstructing the solar magnetic field from Ca II K plage observations.
This difference is most likely due to the high spatial resolution of HMI, since a low-resolution observation averages small elements of opposite polarity as well as weak and strong fields, and thereby leads to smaller observed magnetic field.
The findings of this paper will improve the understanding of the correspondence between FUV emission and magnetic field, the evolution of chromospheric plages, and the accuracy of magnetic field reconstructions based on historical observations of Ca II K spectroheliograms.
They also provide useful information when studying the high-resolution pixel-by-pixel-based reconstructions of the solar magnetic fields and the evolution of solar magnetic fields in flux transport simulations.
However, a detailed study demonstrating that these findings based on AIA 1600~\AA{} observations also apply to Ca II K spectroheliograms will be made in a separate study.

\begin{acknowledgements}
We acknowledge the financial support by the Academy of Finland to the ReSoLVE Centre of Excellence (project no. 307411).
I.T acknowledges the financial support by the Academy of Finland to the PROSPECT (project no. 321440) and by the Finnish Academy of Science and Letters (V{\"a}is{\"a}l{\"a} Fund).
SDO data was downloaded from JSOC data facility courtesy of AIA and HMI science teams.
We thank Wei Liu and AIA team for valuable information on the data calibration.
We thank the anonymous referee for the constructive comments that improved this article.
I.T, I.I.V, A.A.P and K.M are members of international teams on Reconstructing Solar and Heliospheric Magnetic Field Evolution Over the Past Century and Modeling Space Weather And Total Solar Irradiance Over The Past Century supported by the International Space Science Institute (ISSI), Bern, Switzerland and ISSI-Beijing, China.
\end{acknowledgements}

\bibliography{plage_mag_field_bibliography}

\end{document}